\documentclass[review]{elsarticle}

\usepackage{lineno,hyperref}
\usepackage[dvipdfm, usenames]{color}

\journal{Physica A}

\begin{document}

\begin{frontmatter}

\title{Heterogeneity of link weight and the evolution of cooperation}
%% \tnotetext[mytitlenote]{Fully documented templates are available in the elsarticle package on \href{http://www.ctan.org/tex-archive/macros/latex/contrib/elsarticle}{CTAN}.}

\author[mymainaddress]{Manabu Iwata\corref{correspondingauthor}}
\ead{miwata@sk.tsukuba.ac.jp}

\author[mymainaddress]{Eizo Akiyama}
\ead{eizo@sk.tsukuba.ac.jp}

\cortext[correspondingauthor]{Corresponding author at Graduate School of Systems and Information Engineering, University of Tsukuba, 1-1-1 Tennoudai, Tsukuba, Ibaraki 305-0006, Japan. TeL: +81 29 853 5571.}
\address[mymainaddress]{Graduate School of Systems and Information Engineering, University of Tsukuba, 1-1-1 Tennoudai, Tsukuba, Ibaraki 305-0006, Japan}

\begin{abstract}
In this paper, we investigate the effect of \emph{heterogeneity of link weight}, heterogeneity of the frequency or amount of interactions among individuals, on the evolution of cooperation.
Based on an analysis of the evolutionary prisoner's dilemma game on a weighted one-dimensional lattice network with \emph{intra-individual heterogeneity}, we confirm that moderate level of link-weight heterogeneity can facilitate cooperation.
Furthermore, we identify two key mechanisms by which link-weight heterogeneity promotes the evolution of cooperation: mechanisms for spread and maintenance of cooperation.
We also derive the corresponding conditions under which the mechanisms can work through evolutionary dynamics.
\end{abstract}

\begin{keyword}
\texttt{} Evolution of cooperation \sep Prisoner's dilemma \sep Heterogeneity of link weight \sep One-dimensional lattice network \sep Game theory
%% \MSC[2010] 00-01\sep  99-00
\end{keyword}

\end{frontmatter}


\section{Introduction}\label{section: Introduction}

The evolution of cooperation, which plays a key role in natural and social systems, has attracted much interest in diverse academic fields, including biology, sociology, and economics~\cite{Darwin1859, Buss1987}.
The prisoner's dilemma (PD) is often used to study the evolution of cooperation in a population consisting of selfish individuals~\cite{Neumann1953, Axelrod1984}.
In the PD game, two individuals simultaneously decide to cooperate or defect.
A payoff matrix of the PD game is given in Table \ref{Payoff matrix for the PD game}.

\begin{table}[htbp]
\centering
\scalebox{0.9}{
\begin{tabular}{ccc} \hline
& Cooperation & Defection \\ \hline
Cooperation & $R$, $R$ & $S$, $T$ \\
Defection & $T$, $S$ & $P$, $P$ \\ \hline
\end{tabular}
}
\caption{Payoff matrix for the prisoner's dilemma (PD) game.
In this game, two individuals decide simultaneously to cooperate or defect. 
Mutual cooperation provides them both with a payoff $R$, whereas mutual defection results in a payoff $P$.
If one individual cooperates and the other defects, the former obtains a payoff $T$, and the latter a payoff $S$.
These values are assumed to satisfy the conditions $T>R>P>S$ and $2R>S+T$.}
\label{Payoff matrix for the PD game}
\end{table}

If either individual wishes to maximize his/her personal profit in this game, he/she will choose to defect regardless of the opponent's decision, despite mutual cooperation being better than mutual defection for both individuals.
According to the evolutionary dynamics of the PD game where an individual is paired with a randomly chosen opponent in a well-mixed population, cooperators become extinct whereas defectors eventually dominate in the population~\cite{weibull1995}.

However, in a dilemma situation in the real world, we often see that altruistic behaviors exist among unrelated individuals.
Nowak~\cite{Nowak2006} proposed \emph{five rules} as the mechanisms enabling the evolution of altruism: kin selection~\cite{Fisher1930}, direct reciprocity~\cite{Axelrod1984, Trivers1971}, indirect reciprocity~\cite{Alexander1987, Nowak1998}, network reciprocity~\cite{Nowak1992, Nowak1993, Santos2005, Chen2007, Masuda2003, Rong2007, Szabo2007, Perc2010a, Perc2013}, and group selection~\cite{Wynne-Edwards1962}.

In this study, we focus on \emph{network reciprocity}, which is a mechanism pioneered by Nowak and May~\cite{Nowak1992, Nowak1993} that enables the evolution of cooperation when each individual is likely to interact repeatedly with a fixed subset of the population only.
In Nowak and May's model, individuals are placed on nodes in a two-dimensional lattice and play the PD game repeatedly with their directly connected neighbors only.
The authors show that the spatial constraint of interactions among individuals in the lattice network can facilitate the evolution of cooperation.
Although Nowak and May's model assumes that the population has a simple network structure, that is, a two-dimensional lattice, it has recently been shown that many real-world networks are identified as \emph{complex networks}.
Well-known examples of complex networks are the small-world network~\cite{Milgram1967} and the scale-free network~\cite{Barabasi1999}, in which \emph{the number of links} (degree) that each individual has differs.
Recently, it has been confirmed by Santos and Pacheco~\cite{Santos2005} that \emph{heterogeneity of the number of links} in complex networks can enhance the evolution of cooperation.
There have been following studies that investigate the evolution of cooperation on networks with heterogeneous number of links~\cite{Chen2007, Masuda2003}.
This heterogeneity is also known to contribute to the efficiency of collective action~\cite{Liu2014, Gao2015}.
Additionally, it has been shown that the mixing pattern of link degree can affect the emergence of cooperation~\cite{Rong2007}.
See~\cite{Szabo2007, Perc2010a} for detailed reviews of evolutionary and coevolutionary games on graphs.
Also see~\cite{Perc2013} for a thorough survey of the evolutionary dynamics of group interactions on various types of structured populations.

The aforementioned studies, however, assume that individuals interact with one another with the same frequency or amount; that is, all the link weights between individuals in the society are identical.
On the contrary, individuals in real-world networks, such as scientific collaboration networks, phone call networks, email networks, and airport transportation networks, have heterogeneous intentions in their relationships~\cite{Newman2001, Barrat2004, Onnela2007}.
There is substantial interest among researchers in knowing how heterogeneity of the strength of relationships (that is, link weight) among individuals influences human behavioral traits (e.g. sociological studies such as~\cite{Granovetter1973, Bian1997, Yakubovich2005}).

In particular, researchers have recently investigated whether the heterogeneity of link weight between individuals promotes the evolution of cooperation.
For example, Du et al.~\cite{Du2008} constructed a simulation model in which individuals are placed on a node in a scale-free network and connected to other individuals with heterogeneous link weights.
In their model, individuals interact more frequently with neighbors connected by links with large weights and less frequently with those connected by links with small weights.
Du et al. found that cooperative behavior can be more facilitated when the link weights shared by individuals are heterogeneous rather than homogeneous.
Note that, in their model, interaction networks have two kinds of heterogeneity: heterogeneity of the \emph{number of links} and that of \emph{link weight}.
Note also that each link weight is determined according to the number of links of individuals; that is, link weight is a function of the degrees of the two individuals at either side of the focal link.
Therefore, in the Du et al. model it is difficult to ascertain which factor enhances cooperation: \emph{heterogeneity of the number of links} or \emph{heterogeneity of link weight}.

Additionally, Ma et al.~\cite{Ma2011} employed a two-dimensional square lattice with individuals placed on its nodes.
In their model, individuals play the PD game with their immediate neighbors connected by links with heterogeneous link weights.
Ma et al. arranged three populations, where the link weights in the population follow either power-law, exponential, or uniform distribution patterns.
They confirmed that a network with a power-law distribution of link weights better facilitates the evolution of cooperation than one with link weights conforming to one of the other two probability distributions.

Because a two-dimensional square lattice is used in their model, each individual has the same number of links (i.e., four).
Thus, their result clearly shows that \emph{heterogeneity of link weight} can bring about a cooperative state even without \emph{heterogeneity of the number of links}.
However, in their model, the sum of link weights of an individual, which we call the \emph{link-weight amount} of the individual, differs from those of others.
That is, not only each of the links possessed by an individual can have a different weight, but the individual can also have a different link-weight amount from other individuals.
We call the former \emph{intra-individual heterogeneity} and the latter \emph{inter-individual heterogeneity}.

When \emph{inter-individual heterogeneity} exists, some individuals play the PD game more frequently than others (link-weight amount is heterogeneous among individuals).
That is, there is \emph{heterogeneity of the interactions among individuals}.
It has already been shown~\cite{Santos2005, Chen2007, Masuda2003} that heterogeneity of interactions among individuals due to \emph{heterogeneity of the number of links among individuals} and not to \emph{inter-individual heterogeneity} can facilitate the evolution of cooperation.

\begin{figure}[htbp]
\centering
\includegraphics[scale=0.25]{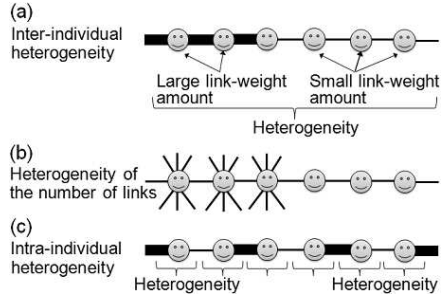}
\caption{Examples of a one-dimensional lattice with three kinds of heterogeneity: (a) \emph{inter-individual heterogeneity}, (b) \emph{heterogeneity of the number of links}, and (c) \emph{intra-individual heterogeneity}.
Thick and thin lines between individuals denote links with large and small weights, respectively.}
\label{Three types of heterogeneity}
\end{figure}

Fig.~\ref{Three types of heterogeneity} shows three examples of a one-dimensional lattice having, respectively, \emph{intra-individual heterogeneity}, \emph{inter-individual heterogeneity}, and \emph{heterogeneity of the number of links}.
Fig.~\ref{Three types of heterogeneity}(a) shows an example of \emph{inter-individual heterogeneity}, where individuals on the left-hand side have large link-weight amounts and those on the right-hand side have small link-weight amounts.
In this case, individuals on the left-hand side interact more frequently with others than those on the right-hand side; that is, there is heterogeneity of interactions between individuals.
\emph{Heterogeneity of the number of links} is shown in Fig.~\ref{Three types of heterogeneity}(b), where individuals on the left-hand side have a large number of links and thus more opportunity to interact than those on the right-hand side.
Both \emph{inter-individual heterogeneity} and \emph{heterogeneity of the number of links} bring about a similar type of heterogeneity of interactions among individuals in the sense that either can cause link-weight amount heterogeneity.
Finally, Fig.~\ref{Three types of heterogeneity}(c) shows an example of \emph{intra-individual heterogeneity}, where each individual has both a large-weight link and a small-weight link, but all individuals have an equivalent link-weight amount.
That is, \emph{intra-individual heterogeneity} does not involve the heterogeneity of the link-weight amount among individuals but involves the heterogeneity of the weight of links of each individual.
Thus, it should be noted that \emph{intra-individual heterogeneity} and \emph{inter-individual heterogeneity} are essentially different types of heterogeneity.

The literature~\cite{Du2008, Ma2011, Du2009, Buesser2011, Cao2011, Buesser2012, Lei2010, Chen2014, Perc2011, Perc2010b, Szolnoki2008, Perc2008, Santos2012, Perc2006a, Perc2006b, Perc2006c, Brede2011, Tanimoto2014} has investigated the effect of heterogeneity on the evolution of cooperation from variety of viewpoints~\footnote{For example, Cao et al.~\cite{Cao2011} focused on the dynamics (change over time) of the magnitude of the link-weight heterogeneity.
Chen and Perc~\cite{Chen2014} examined the effect of the heterogeneity in incentives for rewarding individuals in the public goods game on promoting cooperation.
Szolnoki et al.~\cite{Szolnoki2008}, Perc and Szolnoki~\cite{Perc2008}, and Santos et al.~\cite{Santos2012} examined diversity of individuals.
Perc~\cite{Perc2006a} introduced the random variations to the payoff of individuals and invetigated the effect of it on the evolution of cooperation.
Brede~\cite{Brede2011} and Tanimoto~\cite{Tanimoto2014} analyzed the bias in game partner selection.}.
Especially, Du et al.~\cite{Du2008} and Ma et al.~\cite{Ma2011} have clearly shown, as mentioned in the above, that the existence of link-weight heterogeneity facilitates the evolution of cooperation.
However, we cannot reject the possibility that the evolution of cooperation in the models of Du et al. and Ma et al. might be facilitated by the effect of link-weight amount heterogeneity, whose effect on the evolution of cooperation has already been confirmed by Santos and Pacheco~\cite{Santos2005}.
This is because the link-weight heterogeneity in their models involves not only \emph{intra-individual heterogeneity} but also {\em inter-individual heterogeneity}.
Whether heterogeneity of link weight without \emph{heterogeneity of link-weight amount}, \emph{intra-individual heterogeneity} alone, can promote the evolution of cooperation or not is the remaining question to be solved.
Detailed investigation of this question would enable us to understand the underlying mechanism of the evolution of cooperation caused by the heterogeneity of interactions among individuals.

To answer this question, we introduce the simplest possible model of a weighted network, with \emph{intra-individual heterogeneity} and without \emph{inter-individual heterogeneity}.
First, we employ a weighted one-dimensional lattice as the simplest network model and investigate the effect of link-weight heterogeneity on the enhancement of cooperation.
Second, we examine when and how such heterogeneous link weight gives rise to the evolution of cooperation; that is, we investigate the mechanism by which heterogeneity enables cooperation to evolve.
Specifically, we identify the conditions under which link-weight heterogeneity enables society to become cooperative, through analytical calculation and computer simulation.

\section{The model}\label{section: The model}

In this study, we develop a model of a spatial evolutionary game with link-weight heterogeneity based both on the PD cellular automaton model proposed by Nowak and May~\cite{Nowak1992, Nowak1993} and the weighted network model employed by Du et al.~\cite{Du2008, Du2009}, Ma et al.~\cite{Ma2011}, Buesser~\cite{Buesser2011}, and so on.
We construct a lattice network model in which each individual occupies one node and is connected to his/her neighbors by links with heterogeneous weights.
Each individual has two links, one shared by the individual to his/her left and one to his/her right.
We assume periodic boundary conditions for the network we employ.

There are two types of heterogeneity of link weight in a network: (i) \emph{intra-individual heterogeneity}: the heterogeneity of link weight between the links of an individual; that is, an individual can have a large-weight link with one neighbor and a small-weight link with another; and (i\hspace{-.1em}i) \emph{inter-individual heterogeneity}: the heterogeneity of link weight between individuals; that is, an individual can have many large-weight links whereas another can have many small-weight links. 
In this research, we focus on the former type of heterogeneity to begin our investigation on the effect of link-weight heterogeneity on the evolution of cooperation using the simplest form of heterogeneity.

Because we consider a network with \emph{intra-individual heterogeneity} only (i.e., without \emph{inter-individual heterogeneity}), the sums of the link weights of all the individuals are roughly equivalent.
Let the weight of a link (large-weight link) of an individual be $w_{1}$ and the weight of the other link (small-weight link) be $w_{2}$ ($w_{1}>w_{2}>0$).
Because we assume that there is no \emph{inter-individual heterogeneity} (the sum of $w_{1}$ and $w_{2}$ is the same for all individuals) and that an individual's right (left) link is shared by the right (left) neighbor, all individuals have a link with weight $w_{1}$ and one with weight $w_{2}$.

Hereafter, we assume a large link weight to be $w_{1}=1.0+w$ and a small weight to be $w_{2}=1.0-w$ to express $w_{1}$ and $w_{2}$ using only one parameter, $w\in [0, 1]$.
The larger the value of $w$, the more heterogeneous the link weight becomes.
When $w=0$, link weight in the lattice network is completely homogeneous.

To investigate the evolution of cooperation in weighted networks, we consider the situation where each individual $i$ plays the PD game with his/her immediate neighbors in a weighted lattice network as described above.
We assume an individual $i$ has a strategy $s_{i}=$\{$C, D$\} that determines whether to cooperate with or defect from all of his/her neighbors.
That is, each individual can either be a cooperator who always cooperates with all his/her opponents or a defector who always defects.
According to the literature~\cite{Nowak1992, Nowak1993, Santos2005, Chen2007, Masuda2003, Rong2007, Szabo2007, Perc2010a, Du2008, Ma2011, Du2009, Cao2011, Szolnoki2008, Perc2008, Brede2011}, we rescale the game to be drawn using a single parameter.
For the PD game, we let $T=b$, $R=1$, and $P=S=0$~\footnote{As Nowak and May~\cite{Nowak1992} noted, simulation results are typically not affected by whether the payoff matrix involves $P=S$ or $P>S$, thus, we assume $P=S$ in our study.
This assumption enables the payoff matrix to be expressed and controlled only by one parameter $b$.} to rescale the payoff matrix using one parameter $b$.
This parameter represents the payoff of a defector when exploiting a cooperator, and is constrained by the interval $1.0<b<2.0$.
In each generation, all pairs of connected individuals play the PD game.
After playing the game, each individual obtains the payoff \emph{multiplied by the value of the weight of the link with his/her opponent}.
The weight of the link between individuals $i$ and $j$ is defined as $w_{ij}$, and the payoff of an individual $i$ with strategy $s_{i}$, when playing with an individual $j$ with strategy $s_{j}$, is represented as $\pi_{s_is_j}$.
The total payoff an individual $i$ receives is expressed as $\Pi_{i}=\sum_{j\in V_{i}} \pi_{s_is_j}w_{ij}$, where $V_{i}$ is the set of neighbors of individual $i$.
In the first generation, each individual's strategy, which is either to cooperate or to defect, is randomly determined with a 50 percent probability.
We define the score of each individual in a generation as the sum of the payoffs received from all the games with his/her neighbors.

After all individuals have played the PD game with all their neighbors, each individual imitates the strategy of the individual with the maximum score among all his/her neighbors including him/herself.
An individual does not change his/her strategy if he/she is one of those with the maximum score.
If an individual has more than one neighbor, excluding him/herself, with the maximum score, he/she chooses one of them randomly.
Individuals update their strategies simultaneously, after which one generation is completed.

For the evolutionary simulation, we set the network size $N$=10,000 and set $b$ (the temptation to defect from a cooperator) such that $b\in (1.0, 2.0)$ in steps of 0.01.
We assume that $w\in [0, 1.0]$ in steps of 0.01, and therefore, $1.0+w$ denotes a large weight (strong link) and $1.0-w$ corresponds to a small weight (weak link).

\section{Simulation results and discussion}\label{section: Simulation results and discussion}

To examine how heterogeneity of link weight affects the evolution of cooperation, we analyzed how the degree of link-weight heterogeneity, $w$, affected the resulting \emph{frequency of cooperation} in the population.
The frequency of cooperation in each generation was calculated as the ratio of the number of cooperators to the total population size.
We defined the frequency for a simulation run as the average of the frequency of cooperation over 100 generations after the 2,000th generation. 
We adopted this definition because, although we wished to estimate the frequency at the convergent state, we found that this state sometimes did not converge to a fixed state but went to a periodic state.
(We checked that the frequency of cooperation could reach a steady or periodic state within 2,000 generations.)
We performed 100 runs of the computer simulation for each parameter setting and calculated the average of the frequency of cooperation for all the runs~\footnote{For each parameter setting, all individuals followed the payoff matrix in which parameter $b$ had an identical value and the value of link-weight parameter $w$ was the same for all links.}, which hereafter, we refer to as \emph{the frequency of cooperation}.

\subsection{Overview of the effect of link-weight heterogeneity on the evolution of cooperation}\label{subsection: Overview of the effect of link-weight heterogeneity}

\begin{figure}[htbp]
\centering
\includegraphics[scale=0.22]{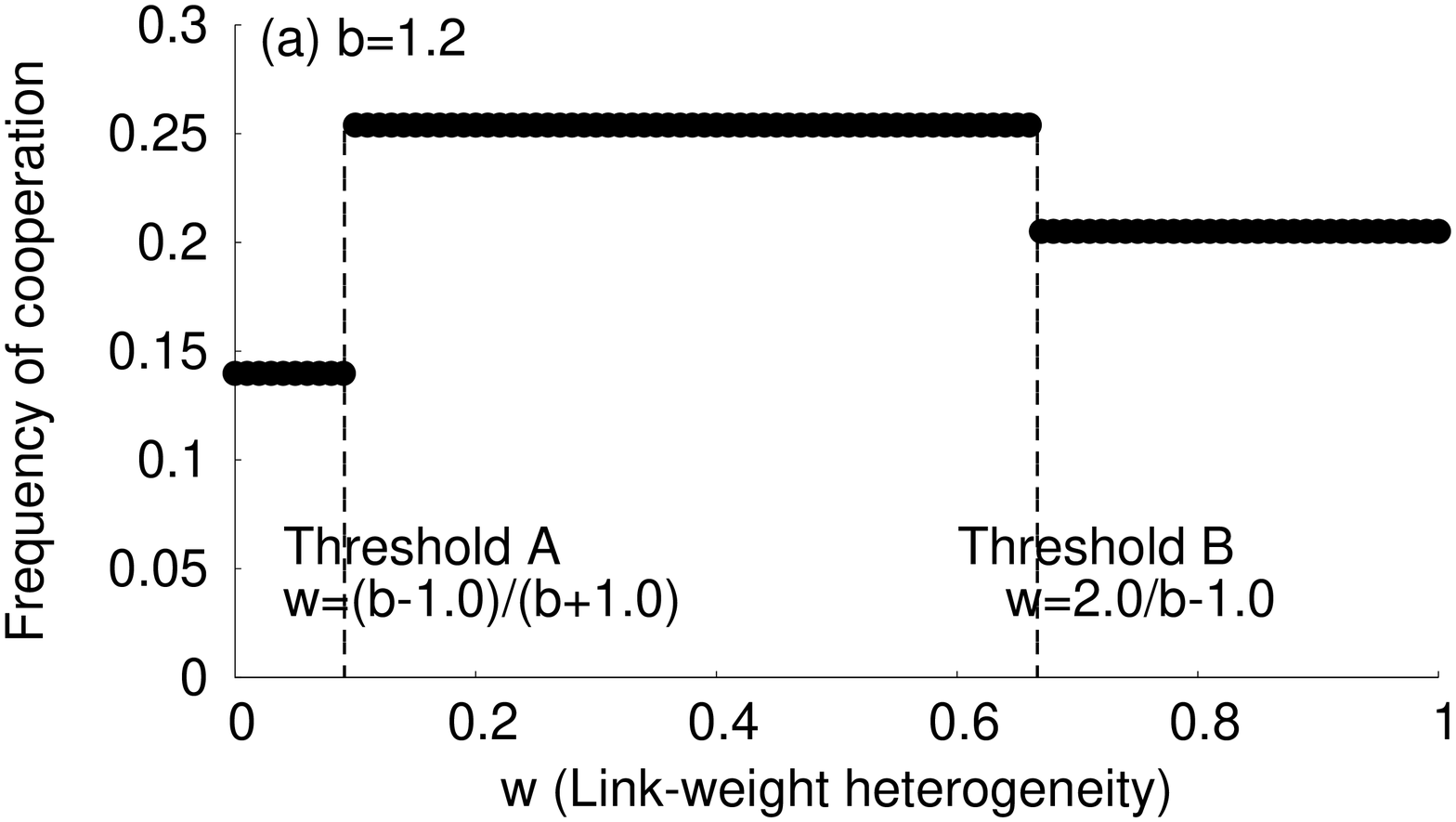}\includegraphics[scale=0.22]{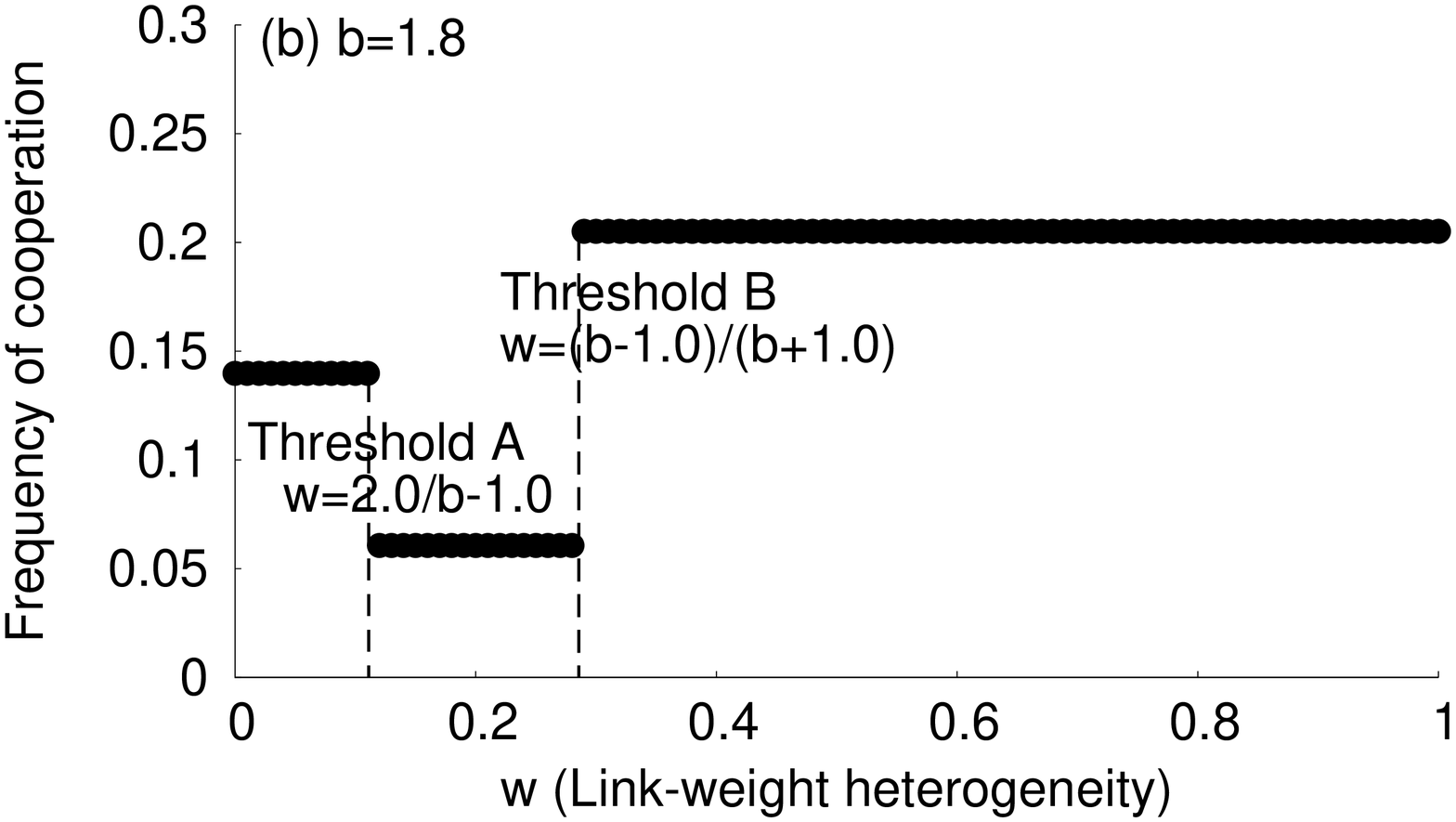}
\caption{Frequency of cooperation for different values of link-weight heterogeneity, $w$, in a weighted one-dimensional lattice: (a) case with a small $b$ ($b$=1.2) and (b) case with a large $b$ ($b$=1.8).
The horizontal and vertical axes represent the degree of $w$, which reflects the magnitude of the heterogeneity of link weight and frequency of cooperation, respectively.}
\label{Cooperation frequency on a one-dimensional lattice}
\end{figure}

Figs.~\ref{Cooperation frequency on a one-dimensional lattice}(a) and (b) illustrate the simulation results for the PD game on a weighted one-dimensional lattice and show
the frequency of cooperation for different values of link-weight heterogeneity, $w$.
As shown in these figures, changes in the frequency of cooperation with an increase in $w$ differ in the case of a small $b=1.2$ (Fig.~\ref{Cooperation frequency on a one-dimensional lattice}(a)) and that of a large $b=1.8$ (Fig.~\ref{Cooperation frequency on a one-dimensional lattice}(b)).

As shown in Fig.~\ref{Cooperation frequency on a one-dimensional lattice}(a), the frequency of cooperation when the link weight is heterogeneous ($w>0$) is always greater than that in the case of homogeneous weight ($w=0$).
Fig.~\ref{Cooperation frequency on a one-dimensional lattice}(b) shows that the frequency of cooperation when the link weight is heterogeneous is smaller than that in the case of homogeneous weight.
If the value of $w$ increases further, however, the magnitude of cooperative behavior in the case of heterogeneous weight is greatly enhanced and exceeds that in the case of homogeneous weight.
In both cases (a) and (b), the cooperation frequency reaches the maximum at some value of $w (>0)$ (when there is some degree of link-weight heterogeneity).
Both of these figures show that the frequency of cooperation does not change with an increase in $w$ until $w$ reaches a certain threshold; that is, the change in the frequency is not gradual, but \emph{stepwise} with an increase in $w$.
As shown in Figs.~\ref{Cooperation frequency on a one-dimensional lattice}(a) and (b), there are two thresholds for $w$, \emph{Threshold A} and \emph{Threshold B}, at which the value of cooperation frequency jumps up or down.
These thresholds are $w=(b-1.0)/(b+1.0)$ and $w=2.0/b-1.0$, the derivations of which are provided later.

We have shown that moderate level of link-weight heterogeneity (intra-individual heterogeneity) can enhance cooperation and that there are some thresholds in $w$ (a parameter that represents the degree of heterogeneity) at which the cooperation frequency changes in a stepwise manner.
We checked that these results are robust against both the difference of the network size and the existence of the decision error (See \emph{Appendix A} for detail).

\subsection{Analysis of small population case --- When and how does the heterogeneity of link weight facilitate cooperation?}\label{subsection: Analysis of small population case}

In the following, we explore why the heterogeneity of link weight brings about the evolution of cooperation and why the frequency of cooperation changes in a stepwise manner with an increase in link-weight heterogeneity $w$.
To answer these questions, we consider a much simpler model composed of six individuals only, and investigate in detail how the heterogeneity of link weight affects evolutionary dynamics.

In the case of a lattice network with six individuals, possible configurations of the strategies chosen by the six individuals are: ``$-$C$\equiv $C$-$C$\equiv $C$-$C$\equiv $C$-$,'' ``$-$C$\equiv $C$-$C$\equiv $C$-$C$\equiv $D$-$,'' ``$-$C$\equiv $C$-$C$\equiv $C$-$D$\equiv $C$-$,'' ..., ``$-$D$\equiv $D$-$D$\equiv $D$-$D$\equiv $D$-$,'' where ``C'' and ``D'' denote cooperator and defector, respectively, ``$\equiv$'' represents a link with a large weight $1.0+w$, and ``$-$'' indicates a link with a small weight $1.0-w$.
The total number of possible configurations is $2^{6}$=64.

Starting with each of the 64 initial configurations, we investigated how the configuration changed over time resulting from updates of the six individuals' strategies and identified the attractors of the evolutionary dynamics of the strategy configurations over generations, which were either steady states or periodic cycles.
Next, we estimated the cooperation frequency in the attractor of evolutionary dynamics by taking an average of the data derived from the last 100 generations.
In each of the 64 case studies, we focused on initial strategy configurations that reached different cooperative states depending on the value of link-weight heterogeneity $w$.
Next, we classified these configurations into three types:
Type (i): configurations that lead to a higher cooperation level when $w>0$ than that with a homogeneous link weight ($w=0$).
Type (i\hspace{-.1em}i): configurations that lead to a lower cooperation level when $w>0$.
Type (i\hspace{-.1em}i\hspace{-.1em}i): configurations leading to the same level of cooperation. (See \emph{Appendix} B for the classification of the initial strategy configurations into these three types.)
Because we are interested in cases where the heterogeneity of link weight has an effect on the evolution of cooperation, in the following, we focus on strategy configurations of the first and second types.

\begin{figure}[htbp]
\centering
\includegraphics[scale=0.2]{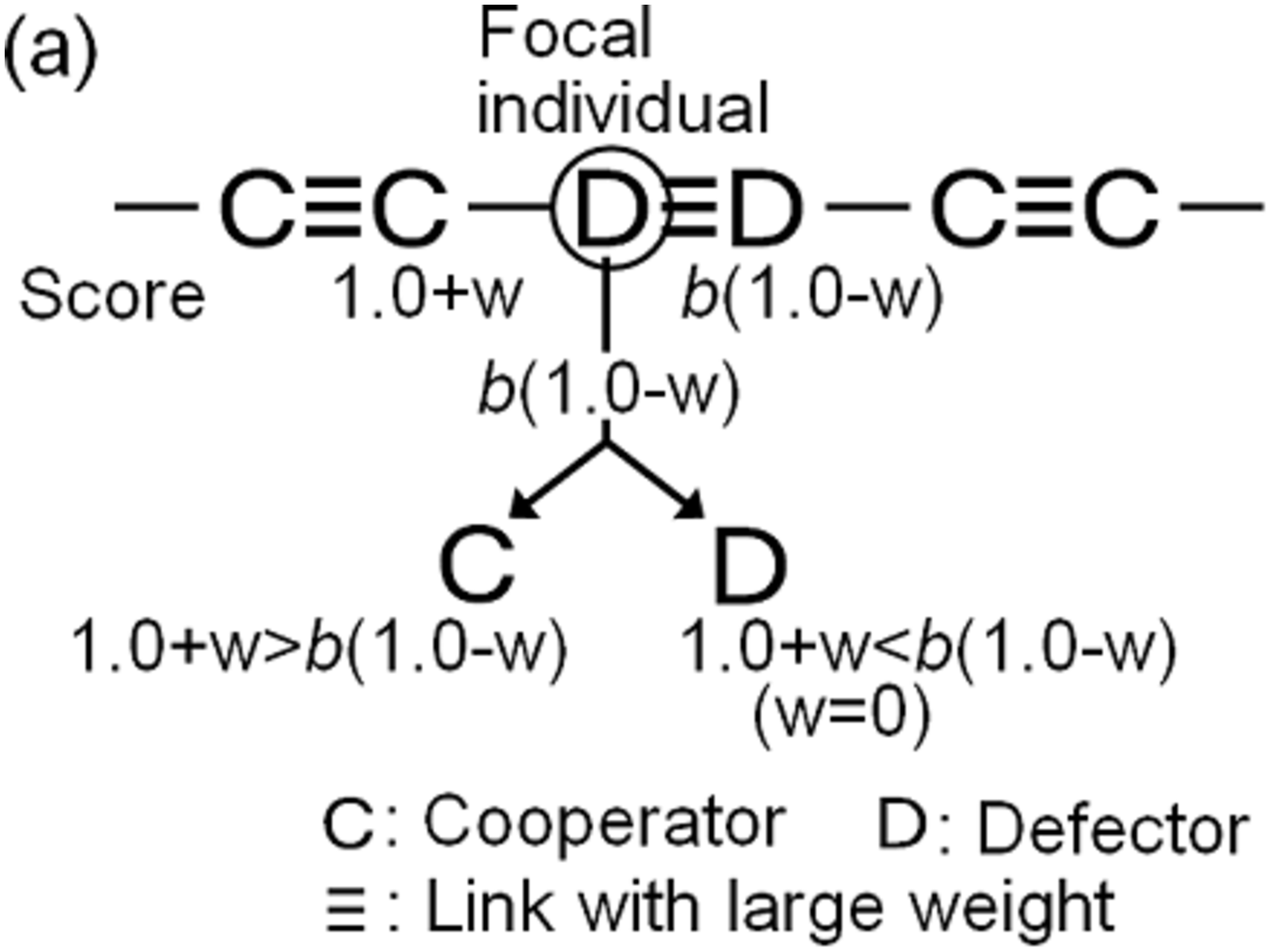}\hspace{7pt}\includegraphics[scale=0.2]{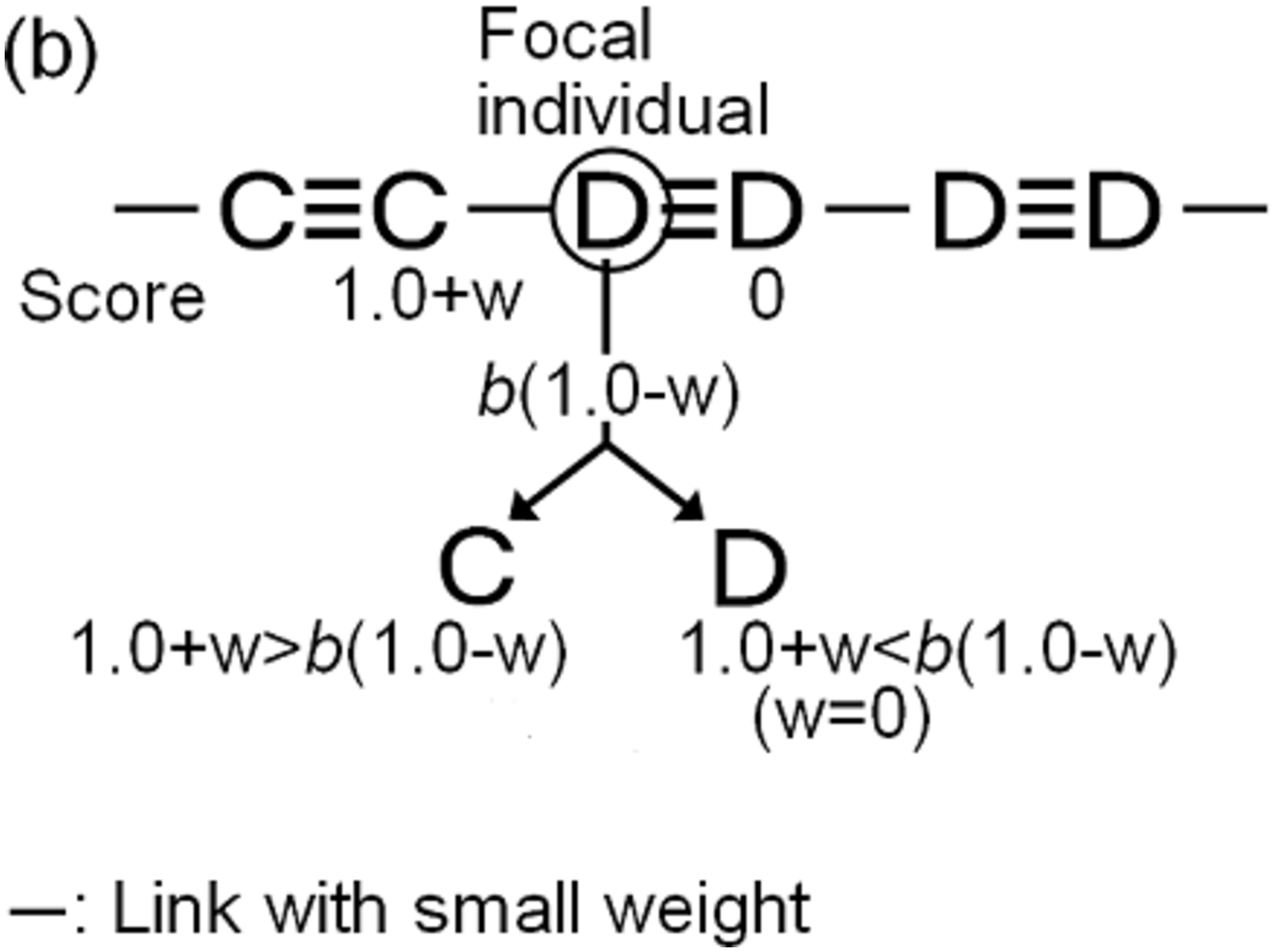}
\caption{Strategy configuration patterns for type (i), where heterogeneous link weight ($w>0$) achieves higher cooperation frequency than the homogeneous one ($w=0$).
``C'' and ``D'' denote cooperator and defector, respectively, ``$\equiv$'' represents a large-weight link, and ``$-$'' indicates a small-weight link.
Whether the strategy of the \emph{focal individual} changes from defection to cooperation depends on the value of link-weight heterogeneity $w$.
The change in strategy of the \emph{focal individual} after one generation (interactions and strategy updates) is indicated by the arrow in the lower part of each figure.}
\label{Strategy configuration patterns: Type(i)}
\end{figure}

\begin{figure}[htbp]
\centering
\includegraphics[scale=0.2]{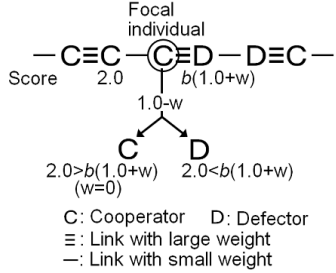}
\caption{Strategy configuration pattern for type (i\hspace{-.1em}i), where heterogeneous link weight ($w>0$) reduces cooperation frequency more than homogeneous link weight ($w=0$).
Here, ``C'' and ``D'' denote cooperator and defector, respectively, ``$\equiv$'' represents a large-weight link, and ``$-$'' indicates a small-weight link.
Whether the \emph{focal individual} changes his/her strategy from cooperation to defection is dependent on the value of link-weight heterogeneity $w$.
A change in the strategy of the \emph{focal individual} after one generation (including interactions and strategy updates) is depicted by the arrow in the lower part of the figure.}
\label{Strategy configuration patterns: Type(ii)}
\end{figure}

There are six strategy configurations that belong to Type (i) (see \emph{Appendix} B), which shows that higher heterogeneity (large $w$) causes an increase in cooperation frequency.
These six configurations have a common pattern, as shown in Figs.~\ref{Strategy configuration patterns: Type(i)}(a) and (b).
Similarly, there are three strategy configurations that belong to Type (i\hspace{-.1em}i), where the magnitude of cooperation frequency decreases with higher heterogeneity.
These three configurations have a common configuration pattern as shown in Fig.~\ref{Strategy configuration patterns: Type(ii)}.

First, we consider the evolutionary dynamics, starting from the strategy configuration pattern in Figs.~\ref{Strategy configuration patterns: Type(i)}(a) and (b).
Here, we focus on the third individual from the left in each of both figures, who we simply call the \emph{focal individual}.
When starting from the strategy configuration pattern shown in these figures, the focal individual chooses cooperation in the next step for a large $w$.
For a small $w$, however, the focal individual does not change his/her strategy and keeps the strategy of defection.
In short, this configuration pattern enables the spread of cooperation if heterogeneity of link weight exists.
We now investigate in detail why this spread of cooperation occurs.

Figs.~\ref{Strategy configuration patterns: Type(i)}(a) and (b) show the strategy configuration patterns for Type (i), where the heterogeneous link weight ($w>0$) achieves a higher cooperation frequency than the homogeneous one ($w=0$).
Of the 64 strategy configurations, there are six configurations where greater link-weight heterogeneity enables a higher cooperation level: ``$-$C$\equiv $C$-$D$\equiv $D$-$C$\equiv $C$-$,'' ``$-$D$\equiv $D$-$C$\equiv $C$-$C$\equiv $C$-$,'' ``$-$C$\equiv $C$-$C$\equiv $C$-$D$\equiv $D$-$,'' ``$-$C$\equiv $C$-$D$\equiv $D$-$D$\equiv $D$-$,'' ``$-$D$\equiv $D$-$D$\equiv $D$-$C$\equiv $C$-$,'' 
and ``$-$D$\equiv $D$-$C$\equiv $C$-$D$\equiv $D$-$.''
Considering the periodic boundary condition, the first three configurations are equivalent to ``$-$C$\equiv $C$-$D$\equiv $D$-$C$\equiv $C$-$,'' which is shown at the top of Fig.~\ref{Strategy configuration patterns: Type(i)}(a). 
Similarly, the latter three configurations are equivalent to ``$-$C$\equiv $C$-$D$\equiv $D$-$D$\equiv $D$-$,'' which is shown at the top of Fig.~\ref{Strategy configuration patterns: Type(i)}(b).
We define two cooperators connected by a large-weight link as ``C$\equiv $C cluster,'' two defectors connected by a large-weight link as ``D$\equiv $D cluster,'' and one cooperator and one defector connected strongly as ``C$\equiv $D cluster'' or ``D$\equiv $C cluster.''
When ``C$\equiv $C cluster,'' ``D$\equiv $D cluster,'' and ``C$\equiv $C or D$\equiv $D cluster'' are adjacent as shown in Figs.~\ref{Strategy configuration patterns: Type(i)}(a) and (b), there is the possibility that the focal individual (the third individual) updates his/her strategy from defection to cooperation depending on the value of $w$ (link-weight heterogeneity).
We call this configuration pattern the \emph{spread pattern strategy configuration}.

Given that there exists a \emph{spread pattern strategy configuration}, we investigate the actual conditions under which the focal individual (the third individual) updates his/her strategy from defection to cooperation.
Because each individual obtains the payoff of the PD game \emph{multiplied by the value of the weight of the link with his/her opponent} and each individual's score is the sum of all the payoffs obtained by playing PD games with his/her neighbors, the score of the third individual is $b(1.0-w)$.
The fourth individual obtains a score of $b(1.0-w)$ if the fifth individual is a cooperator (see Fig.~\ref{Strategy configuration patterns: Type(i)}(a)), else 0 if the fifth is a defector (see Fig.~\ref{Strategy configuration patterns: Type(i)}(b)).
Thus, in either case, the score of the focal individual (the third individual) is greater than or equal to that of the fourth individual.
Because the focal individual is assumed to imitate the strategy of the individual with the maximum score, it is sufficient for the focal individual to compare his/her score with that of the second individual to ascertain whose strategy to imitate.
The focal individual imitates the second individual's strategy and changes his/her strategy from defection to cooperation only if the second individual's score is higher than the focal individual's own score.
Because the score of the focal individual is $b(1.0-w)$ and that of the second individual is $1.0+w$, the condition under which the focal individual imitates the strategy of the second individual is $1.0+w>b(1.0-w)$; that is, $w>(b-1.0)/(b+1.0)$ for a given $b$.

Thus, if the population involves the \emph{spread pattern strategy configuration} composed of three adjoining clusters, namely, ``C$\equiv $C cluster,'' ``D$\equiv $D cluster,'' and ``C$\equiv $C or D$\equiv $D cluster,'' whether the focal individual changes his/her strategy from defection to cooperation depends on the link-weight heterogeneity; that is, he/she becomes a cooperator if the link weight satisfies the condition $w>(b-1.0)/(b+1.0)$.
We refer to the inequality of $w$ mentioned above as the \emph{condition for the spread of cooperation}. 
To summarize, if this condition is satisfied in the \emph{spread pattern strategy configuration}, cooperative behavior spreads from the second to the third individual.

Next, we look at the evolutionary dynamics starting from the strategy configuration pattern in Fig.~\ref{Strategy configuration patterns: Type(ii)}. 
As in the case of Figs.~\ref{Strategy configuration patterns: Type(i)}(a) and (b), we focus on the third individual from the left in this figure and call him/her the \emph{focal individual}. 
When starting from the strategy configuration pattern shown in Fig.~\ref{Strategy configuration patterns: Type(ii)}, the focal individual chooses defection in the next step for sufficiently large values of $w$. 
For a small $w$, however, the focal individual does not change his/her strategy and retains a cooperative state. 
In short, this configuration pattern enables maintenance of cooperation when the heterogeneity is not so large. 
In the following, we investigate in detail why this maintenance of cooperation occurs. 

Fig.~\ref{Strategy configuration patterns: Type(ii)} shows the strategy configuration pattern in the case of Type (i\hspace{-.1em}i), where homogeneous link weight ($w=0$) achieves higher cooperation frequency than heterogeneous weight ($w>0$).
Of the 64 strategy configurations, there are three configurations where a small heterogeneity promotes further cooperation: ``$-$C$\equiv $C$-$C$\equiv $D$-$D$\equiv $C$-$,'' ``$-$C$\equiv $D$-$D$\equiv $C$-$C$\equiv $C$-$,'' and ``$-$D$\equiv $C$-$C$\equiv $C$-$C$\equiv $D$-$.''
Considering the periodic boundary condition, these configurations are equivalent to ``$-$C$\equiv $C$-$C$\equiv $D$-$D$\equiv $C$-$,'' as shown at the top of Fig.~\ref{Strategy configuration patterns: Type(ii)}. 
When ``C$\equiv $C cluster,'' ``C$\equiv $D cluster,'' and ``D$\equiv $C cluster'' are adjacent, as shown in this figure, there is a possibility that the focal individual (third individual) updates his/her strategy from cooperation to defection depending on the value of $w$ (link-weight heterogeneity).
We call this configuration pattern the \emph{maintenance pattern strategy configuration}.

Given that there exists a \emph{maintenance pattern strategy configuration}, we investigate the condition under which the focal individual (third individual) does not update his/her strategy from cooperation to defection and retains the strategy of cooperation.
In this case, the focal individual (third individual) has a score of $1.0-w$, the second has a score of $2.0$, and the fourth has a score of $b(1.0+w)$.
So the focal individual does not imitate the fourth individual's strategy (defection) but imitates the second individual's strategy (cooperation), only if $2.0>b(1.0+w)$; that is, $w<2.0/b-1.0$ for a given $b$.

Therefore, if the population is subject to a \emph{maintenance pattern strategy configuration} composed of three adjoining clusters, namely, ``C$\equiv $C cluster,'' ``C$\equiv $D cluster,'' and ``D$\equiv $C cluster,'' whether the focal individual can refrain from changing his/her strategy from cooperation to defection depends on link-weight heterogeneity; that is, he/she remains a cooperator if the heterogeneity $w$ satisfies the condition $w<2.0/b-1.0$.
We call the inequality of $w$ given above the \emph{condition for maintenance of cooperation}.
If this condition is satisfied in the \emph{maintenance pattern strategy configuration}, defective behavior does not spread from the fourth to the third individual.
Otherwise, defection spreads.

\begin{figure}[htbp]
\centering
\includegraphics[scale=0.22]{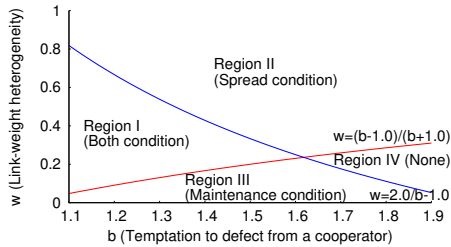}
\caption{Two conditions under which link-weight heterogeneity enables the spread/maintenance of cooperation in a small population.
These conditions are determined and illustrated here using a combination of link-weight heterogeneity $w$ and payoff $b$.
The horizontal and vertical axes represent the payoff $b$ and the value of $w$.}
\label{Two conditions in the small population}
\end{figure}

Thus far, we have derived the condition for the spread of cooperation $w>(b-1.0)/(b+1.0)$\footnote{In fact, the condition $w>(b-1.0)/(b+1.0)$ is not only the spread condition of cooperation but also is the maintenance condition, under which cooperator avoids from changing his/her strategy to defection; that is, cooperation can be maintained.
However, this fact does not change our results in which the intermediate level of the magnitude of heterogeneity can enhance cooperation and that heterogeneity has several thresholds at which cooperation frequency changes in a stepwise manner.
Thus, we omit the fact that $w>(b-1.0)/(b+1.0)$ involves both spread and maintenance conditions in this paper.} and that for the maintenance of cooperation $w<2.0/b-1.0$ for one individual in a small population.
These obtained conditions are illustrated in Fig.~\ref{Two conditions in the small population}.

The parameter space for the temptation payoff, $b$, and link-weight heterogeneity, $w$, is divided into four regions, namely, region I, where both conditions are satisfied, region II, where only the spread condition is satisfied, region III where only the maintenance condition is satisfied, and region IV where neither condition is satisfied.

\subsection{Simulation analysis on a large population}\label{subsection: Simulation analysis on a large population}

The two conditions identified in the previous subsection are based on the analysis of a small population (six nodes); nevertheless, whether these two conditions can control the spread/maintenance of the frequency of cooperation in a large population as well, remains to be seen.
As mentioned, Figs.~\ref{Cooperation frequency on a one-dimensional lattice}(a) and (b) depict the simulation results for a large (10,000 node) one-dimensional lattice showing how link-weight heterogeneity $w$ affects the frequency of cooperation.
By comparing Figs.~\ref{Cooperation frequency on a one-dimensional lattice}(a) and (b) with Fig.~\ref{Two conditions in the small population}, we can see whether the condition for the spread of cooperation $w>(b-1.0)/(b+1.0)$ and that for the maintenance of cooperation $w<2.0/b-1.0$ identified in the small population, also hold in a large population.

For example, when $b=1.2$, the phase shifts in Fig.~\ref{Two conditions in the small population} through regions III (maintenance condition holds), I (both conditions hold), and II (spread condition holds), as parameter $w$ increases.
After an increase in $w$, it will be on the boundary between regions III and I where $w=(b-1.0)/(b+1.0)$ holds. 
As mentioned, in Fig.~\ref{Cooperation frequency on a one-dimensional lattice}(a), if $w$ has a value satisfying $w=(b-1.0)/(b+1.0)$, $w$ is at \emph{Threshold A}, at which point the cooperation frequency increases in a stepwise manner.
If the value of $w$ satisfies $(b-1.0)/(b+1.0)<w<2.0/b-1.0$, cooperation frequency is at its highest value in Fig.~\ref{Cooperation frequency on a one-dimensional lattice}(a), and when $b=1.2$, $(w, b)$ is in region I in Fig.~\ref{Two conditions in the small population}.
Thereafter, if the value of $w$ is equal to $w=2.0/b-1.0$, $w$ is at \emph{Threshold B} at which point the cooperation frequency starts to decrease in a stepwise manner in Fig.~\ref{Cooperation frequency on a one-dimensional lattice}(a), and $(w, b)$ is located at the boundary between regions I and II in Fig.~\ref{Two conditions in the small population}.
Similarly, also in the case where $b=1.8$, the phase shifts in Fig.~\ref{Two conditions in the small population} through regions as an increase in the value of $w$ correspond to the changes in the cooperation frequency seen in Fig.~\ref{Cooperation frequency on a one-dimensional lattice}(b).

We have found that the two conditions identified in the small group can explain the effect of link-weight heterogeneity on the level of cooperation frequency in a large population.
However, so far we have confirmed this only in the cases with $b=1.2$ and $b=1.8$.
Next, we examine the possibility of an application of the two obtained conditions for $w$ to an increase or decrease in the cooperation frequency in a stepwise manner for several thresholds of $w$ over the whole parameter range of $b\in (1.0, 2.0)$ (in steps of 0.01).

\begin{figure}[htbp]
\centering
\includegraphics[scale=0.32]{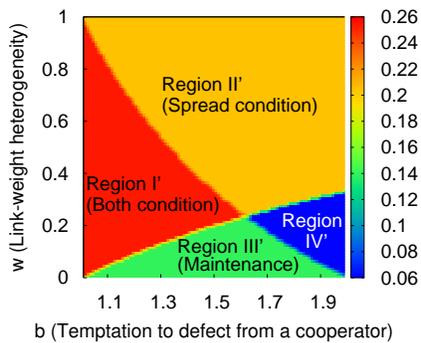}
\caption{Simulation results for the PD game showing the relationship between the frequency of cooperation and $b-w$ parameter combination.
The horizontal and vertical axes denote temptation payoff $b$ and link-weight heterogeneity $w$.
Here, color coding represents the magnitude of the frequency of cooperation, as shown on the sidebar.
We denote the region with the highest cooperation frequency (red) as region I', that with the second highest frequency (orange) as region II', that with the third highest (green) as region III', and the lowest frequency region (blue) as region IV'.}
\label{Two conditions in the large population}
\end{figure}

Fig.~\ref{Two conditions in the large population} illustrates the frequency of cooperation for different values of the combination of $b$ and $w$.
If the combination of $b$ and $w$ denotes a point that is located in region I (both conditions hold) in Fig.~\ref{Two conditions in the small population}, the same point is placed in region I' in Fig.~\ref{Two conditions in the large population}, at which point the cooperation frequency has its highest value.
In addition, if the combination of $b$ and $w$ denotes a point that is located in region II (spread condition holds) in Fig.~\ref{Two conditions in the small population}, the same point is placed in region II' in Fig.~\ref{Two conditions in the large population} and the magnitude of the cooperation frequency is the second highest.
It is observed that the two lines $w=(b-1.0)/(b+1.0)$ and $w=2.0/b-1.0$ in Fig.~\ref{Two conditions in the small population} coincide with the lines dividing the parameter space into four regions (region I', II', III'', and IV') in Fig.~\ref{Two conditions in the large population}.
This coincidence implies that the two conditions for link-weight heterogeneity $w$ identified in the small population also hold for the spread/maintenance of cooperation in the large population across the entire parameter range of $b$.

\section{Conclusion}\label{section: Conclusion}

Much research has been conducted to analyze the factors that promote the evolution of cooperation in natural and social systems.
Recently, several researchers~\cite{Du2008, Ma2011, Du2009, Buesser2011, Cao2011, Buesser2012, Lei2010, Chen2014, Perc2011, Perc2010b, Szolnoki2008, Perc2008, Santos2012, Perc2006a, Perc2006b, Perc2006c, Brede2011, Tanimoto2014} have examined the effect of heterogeneity on the evolution of cooperation.
Especially, Du et al.~\cite{Du2008} and Ma et al.~\cite{Ma2011} have clarified that link-weight heterogeneity can facilitate cooperation.
However, they investigated heterogeneity of interactions among individuals, which includes both \emph{intra-individual heterogeneity} and \emph{inter-individual heterogeneity}, to the best of our knowledge.
\emph{Inter-individual heterogeneity} leads to heterogeneity of the \emph{link-weight amount}, which causes heterogeneous interactions similar to those caused by the \emph{heterogeneity of the number of links}, and the effect of the heterogeneity of the number of links on the promotion of cooperation has already been established in the literature~\cite{Santos2005, Chen2007, Masuda2003}.
Therefore, the effect of link-weight heterogeneity on the evolution of cooperation may be given only by \emph{inter-individual heterogeneity} whose effect is similar to that of the \emph{heterogeneity of the number of links}.
To investigate whether link-weight heterogeneity within each individual alone can promote cooperation, it is necessary to use a model with \emph{intra-individual heterogeneity} and without \emph{inter-individual heterogeneity}.
Additionally, it has not been fully resolved \emph{when and how} promotion of cooperation based on the heterogeneity of link weight takes place.

To address these issues, we constructed a simple model of one-dimensional lattice with heterogeneous link weight, on which individuals play the evolutionary PD game.
We assumed that the sum of the link weights of each individual was equal, to remove the effect of \emph{inter-individual heterogeneity} on the promotion of cooperation, thereby focusing only on \emph{intra-individual heterogeneity}.

By performing calculations and analyses, we obtained the following two results.
First, we clarified that the moderate magnitude of \emph{intra-individual heterogeneity} of link weight can facilitate cooperation and that there are some thresholds in the range of the heterogeneity level, at which the change in the cooperation frequency occurs in a stepwise manner.
This result suggests that, even when there is no heterogeneity of link-weight amount that causes a similar effect to that of heterogeneity of the number of links as in Santos and Pacheco~\cite{Santos2005}, heterogeneous link weight within each individual alone can promote cooperation.
Second, we found the key mechanisms whereby link-weight heterogeneity facilitates the evolution of cooperation, the mechanisms for the spread and maintenance of cooperation.
We also derived corresponding conditions for the both mechanisms to work through evolutionary dynamics, which have not been clarified before.

Because the simulation model used is very simple, it may appear to be somewhat unrealistic.
However, this simplicity enabled us to examine the effect of heterogeneous link weight (\emph{intra-individual heterogeneity}) and the aforementioned mechanisms.
We believe that our discovery of these mechanisms can form the basis of future researches on link-weight heterogeneity.
It would be interesting to investigate the effect of heterogeneity of link weight on the evolution of cooperation and its mechanism using a mathematical model with a more realistic assumption.
For example, to extend the network structure from a one-dimensional lattice to a two-dimensional one~\footnote{Preliminary analyses on a two-dimensional lattice are given in \emph{Appendix C}.}, or so-called complex networks such as small-world and scale-free networks, would be attractive matter to be worked on as a future work.
Another interesting avenue for future research would be to identify the mechanisms by which link-weight heterogeneity that includes both \emph{intra-individual heterogeneity} and \emph{inter-individual heterogeneity}, such as link-weight heterogeneity in the real world, promotes cooperation.
Although we found in this paper the mechanism by which \emph{intra-individual heterogeneity} alone can facilitate cooperation, there may be a specific mechanism for the evolution of cooperation caused by the interplay between \emph{intra-} and \emph{inter-individual heterogeneities}.

\section*{Acknowledgements}\label{section: Acknowledgements}
This work was supported by JSPS KAKENHI Grant Numbers 26350415, 26245026, 26289170, 25242029.

\section*{References}\label{section: References}
\bibliography{mybibfile}

\newpage
\appendix

\section{Investigation of the robustness of the results}

In section \ref{subsection: Overview of the effect of link-weight heterogeneity}, we described that moderate level of link-weight heterogeneity (intra-individual heterogeneity) can promote more cooperation than in the case where link-weight is homogeneous.
Moreover, there are some thresholds in $w$ (a parameter that represents the degree of heterogeneity) at which the stepwise changes of cooperation frequency occur.
To confirm the robustness of these results, we examined whether the difference of the network size and the existence of the decision error affect the evolution of cooperation.

\subsection{Robustness of the results against the system size}

To check the robustness of these results for variations in system size, we employed six one-dimensional lattice networks composed of 100, 500, 1,000, 5,000, and 10,000 individuals.
By performing the computer simulation provided in section 2, we calculated the cooperation frequency and compared it with that derived from the model with the network size of 10,000 individuals.

\setcounter{figure}{0}
\begin{figure}[htbp]
%%\centering
\includegraphics[scale=0.31]{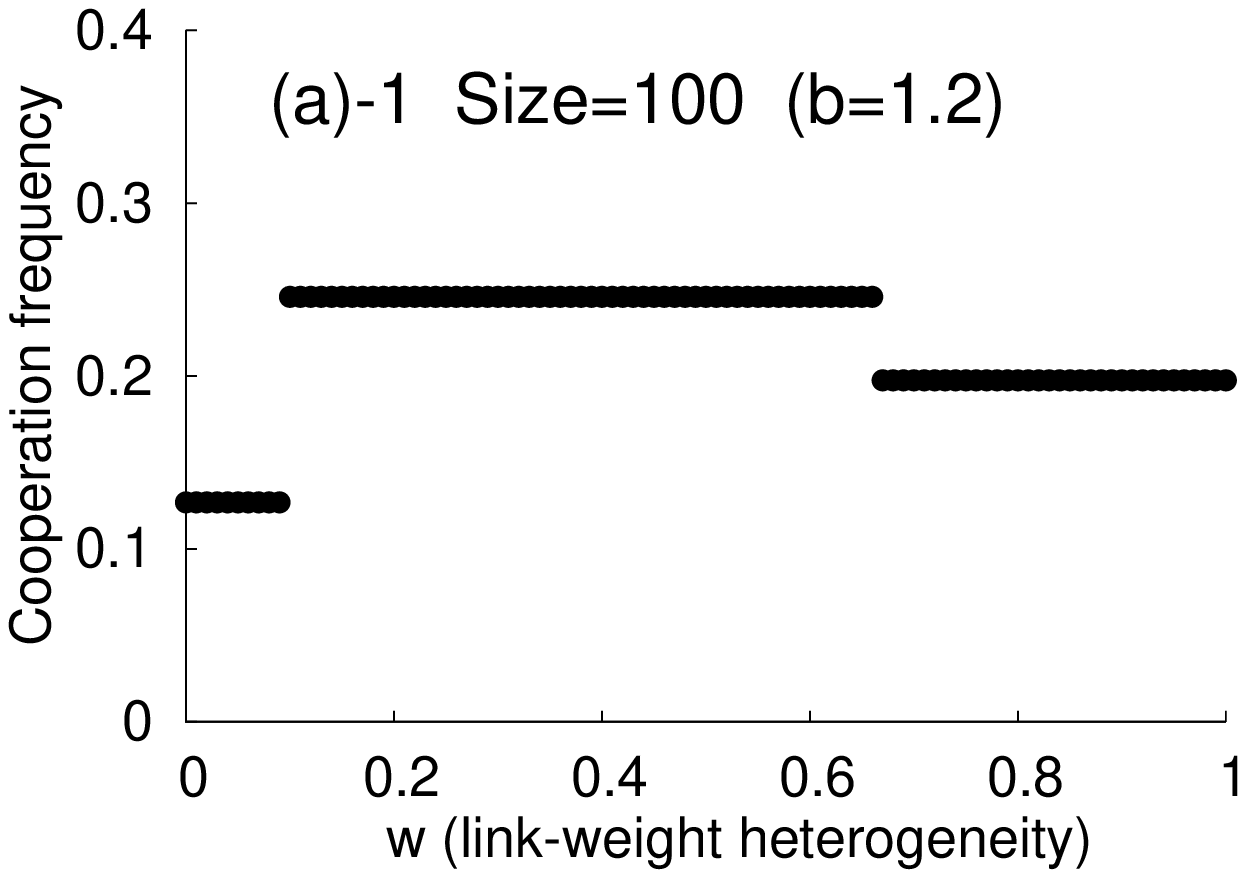}\hspace{3.5pt}\includegraphics[scale=0.31]{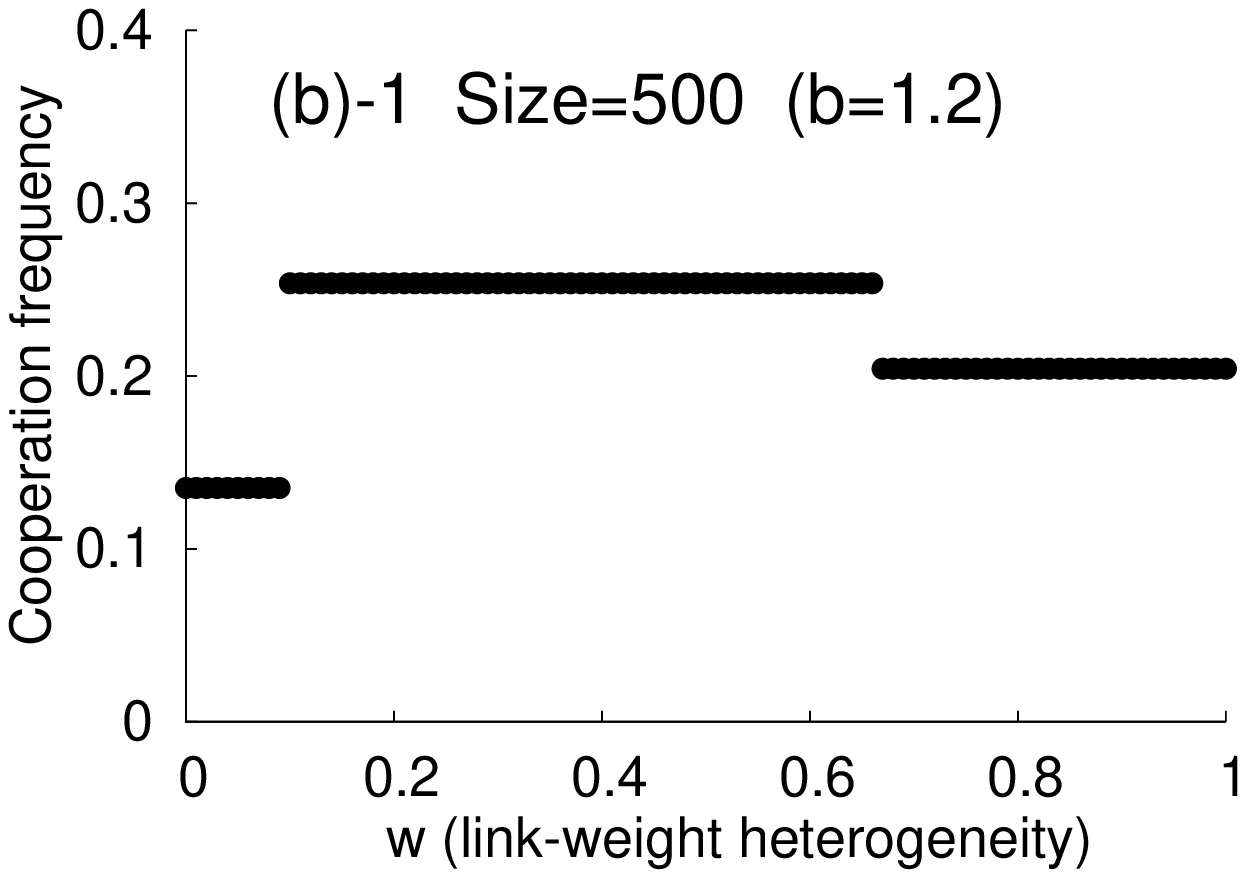}\hspace{3.5pt}\includegraphics[scale=0.31]{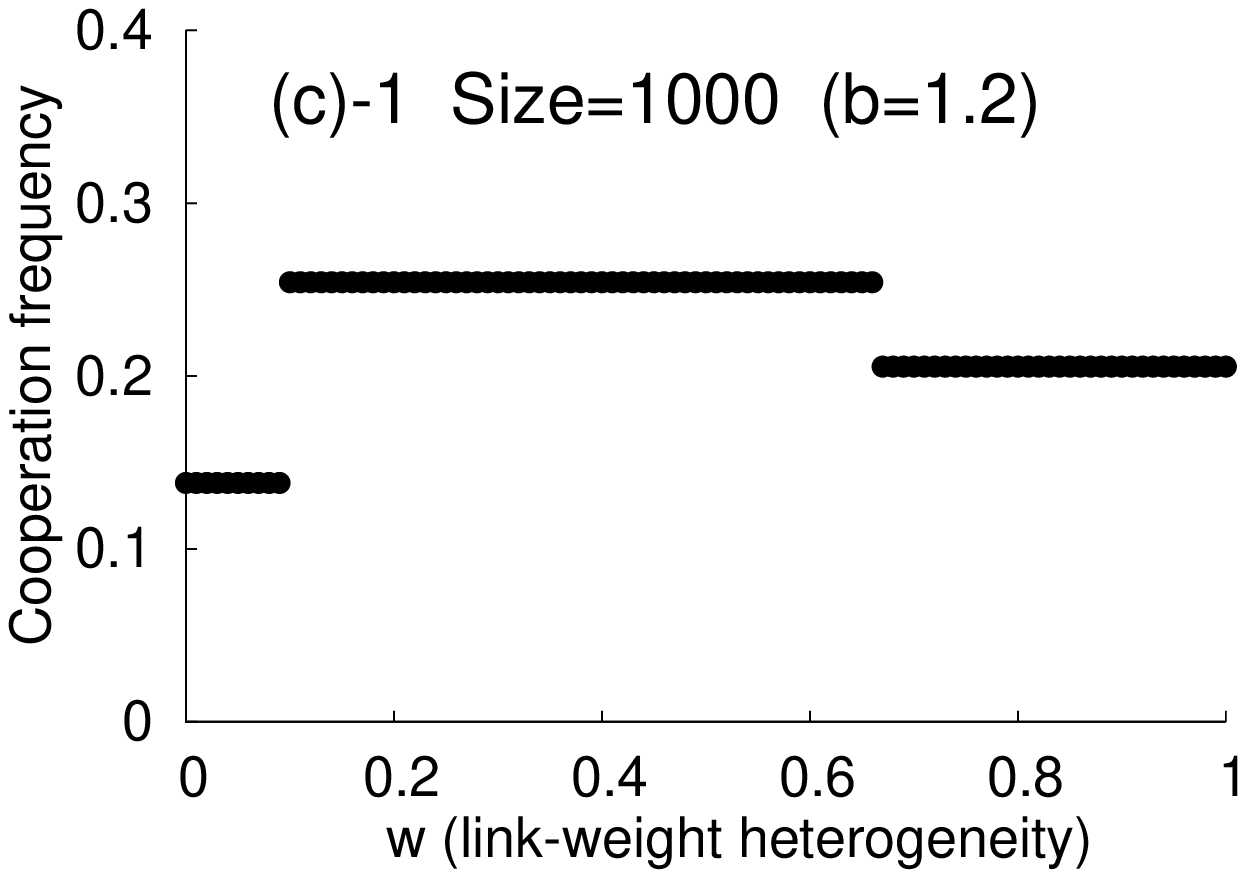}
\includegraphics[scale=0.31]{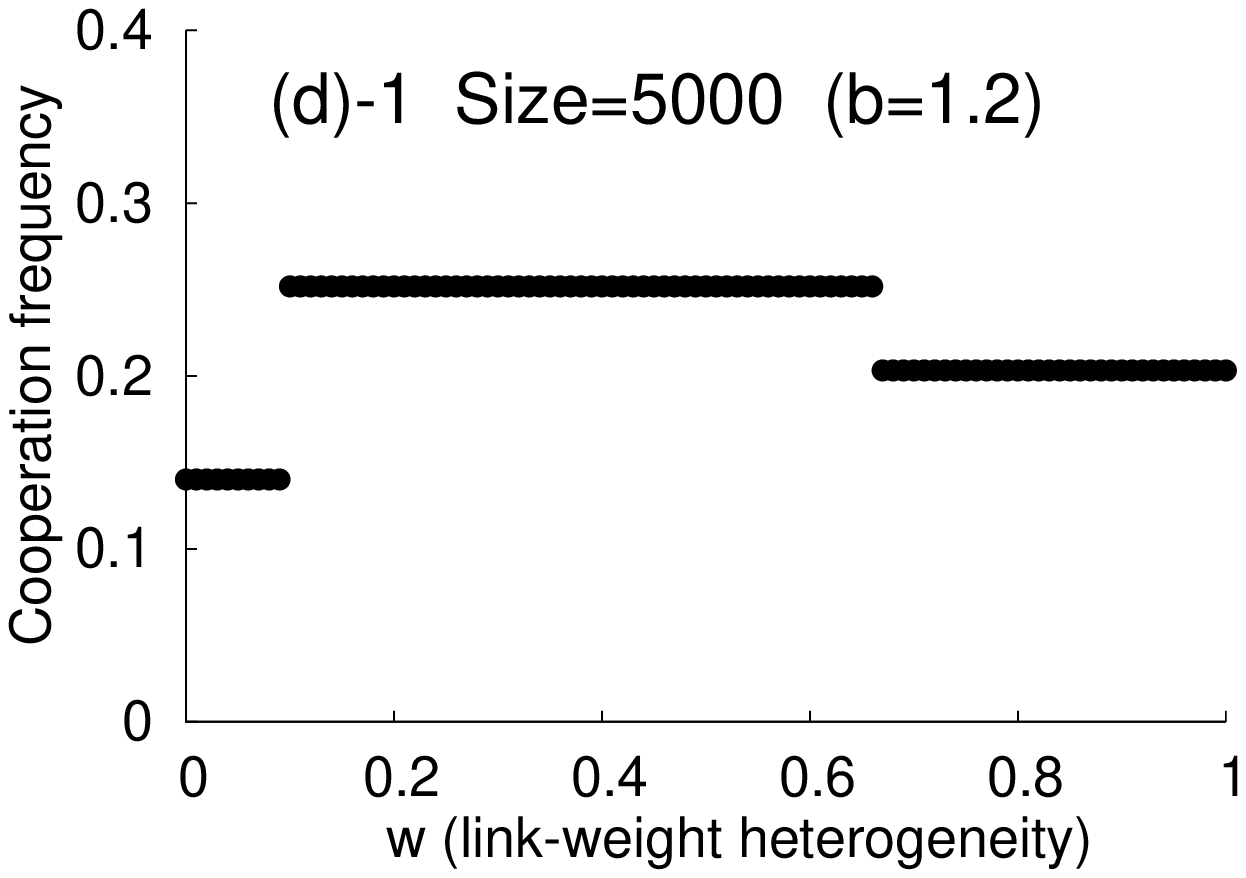}\hspace{3.5pt}\includegraphics[scale=0.31]{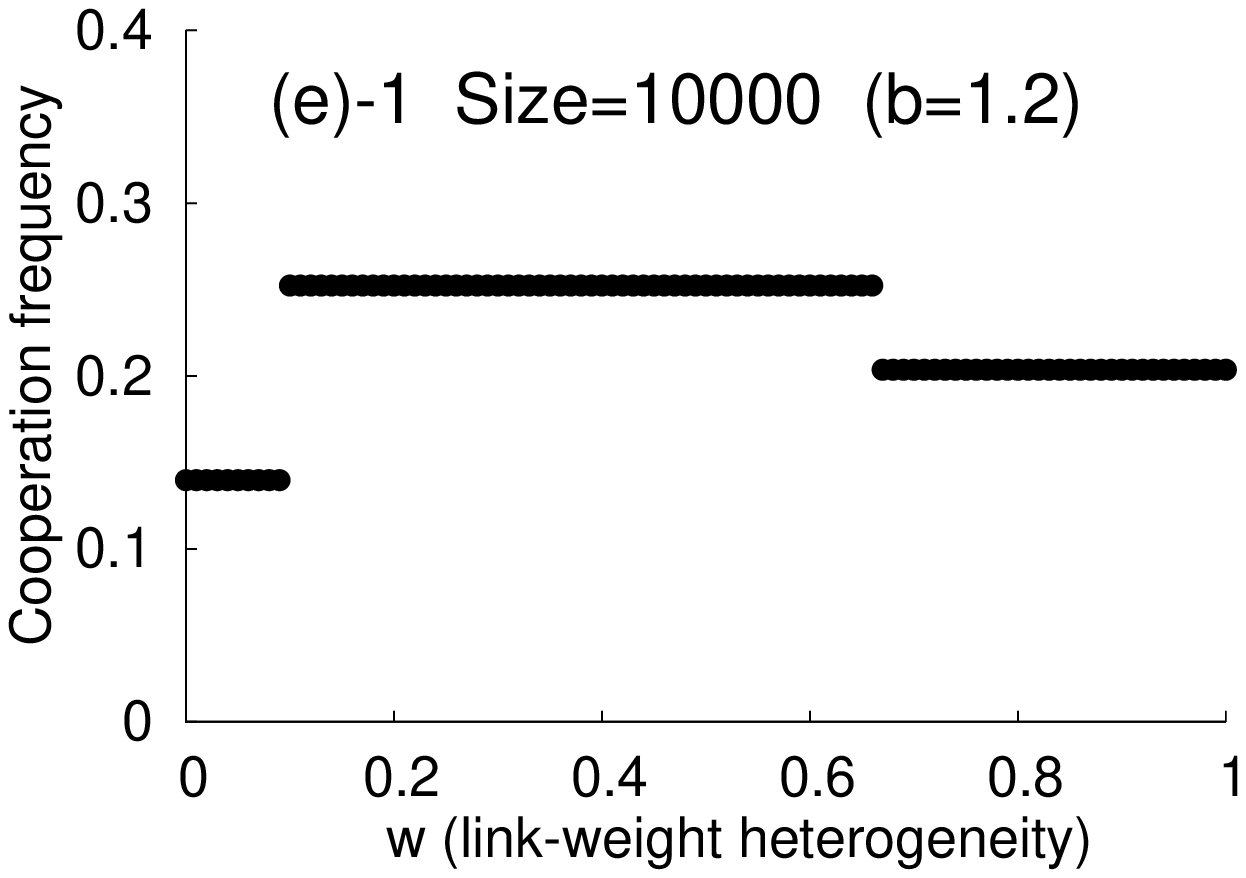}

\includegraphics[scale=0.31]{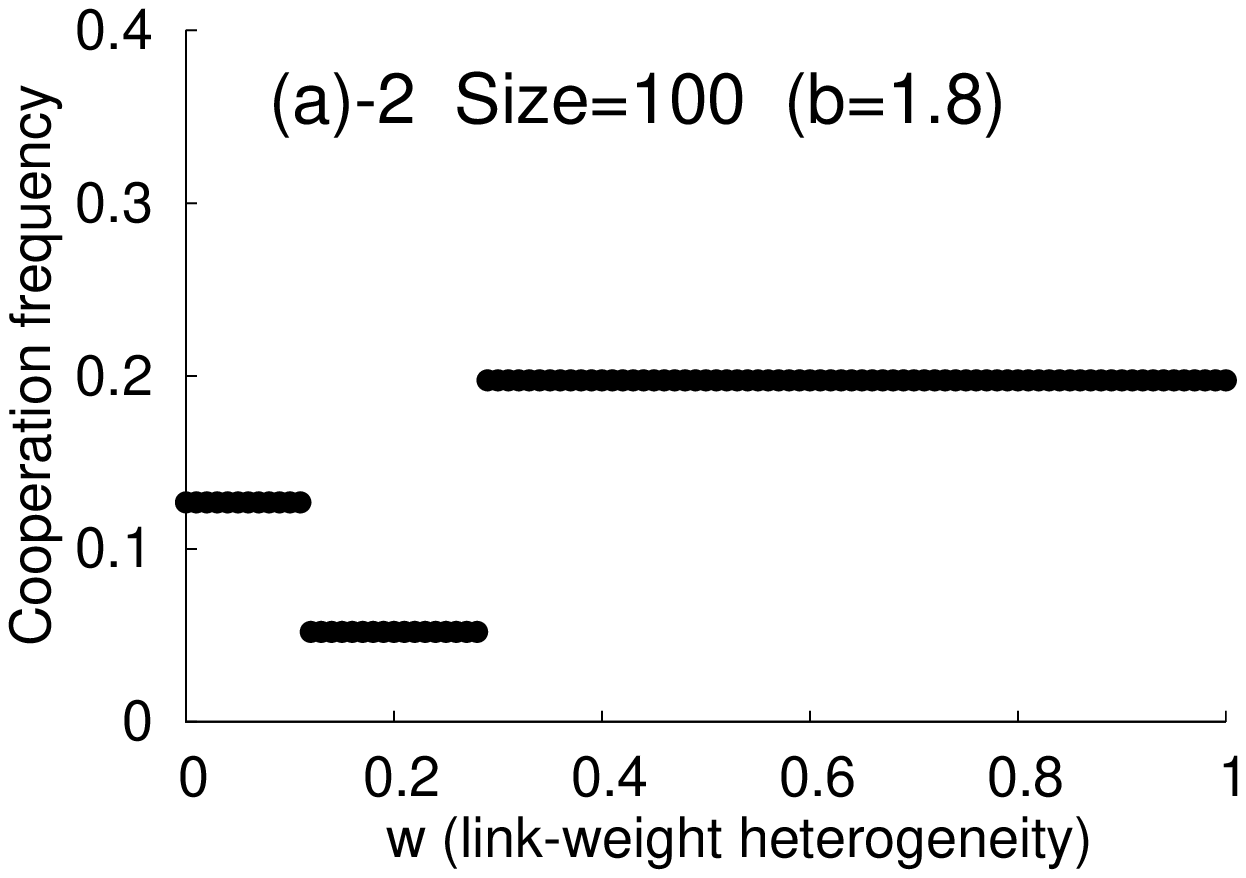}\hspace{3.5pt}\includegraphics[scale=0.31]{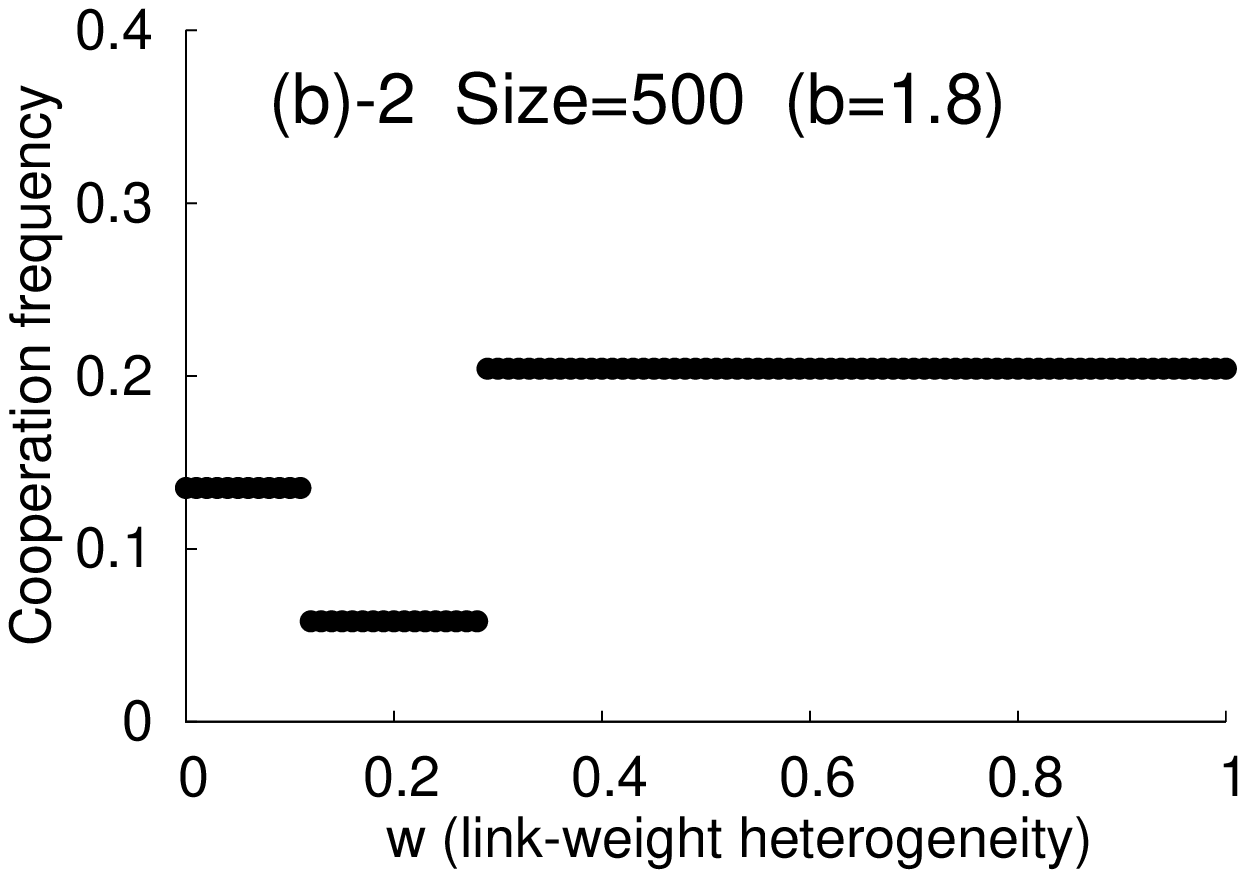}\hspace{3.5pt}\includegraphics[scale=0.31]{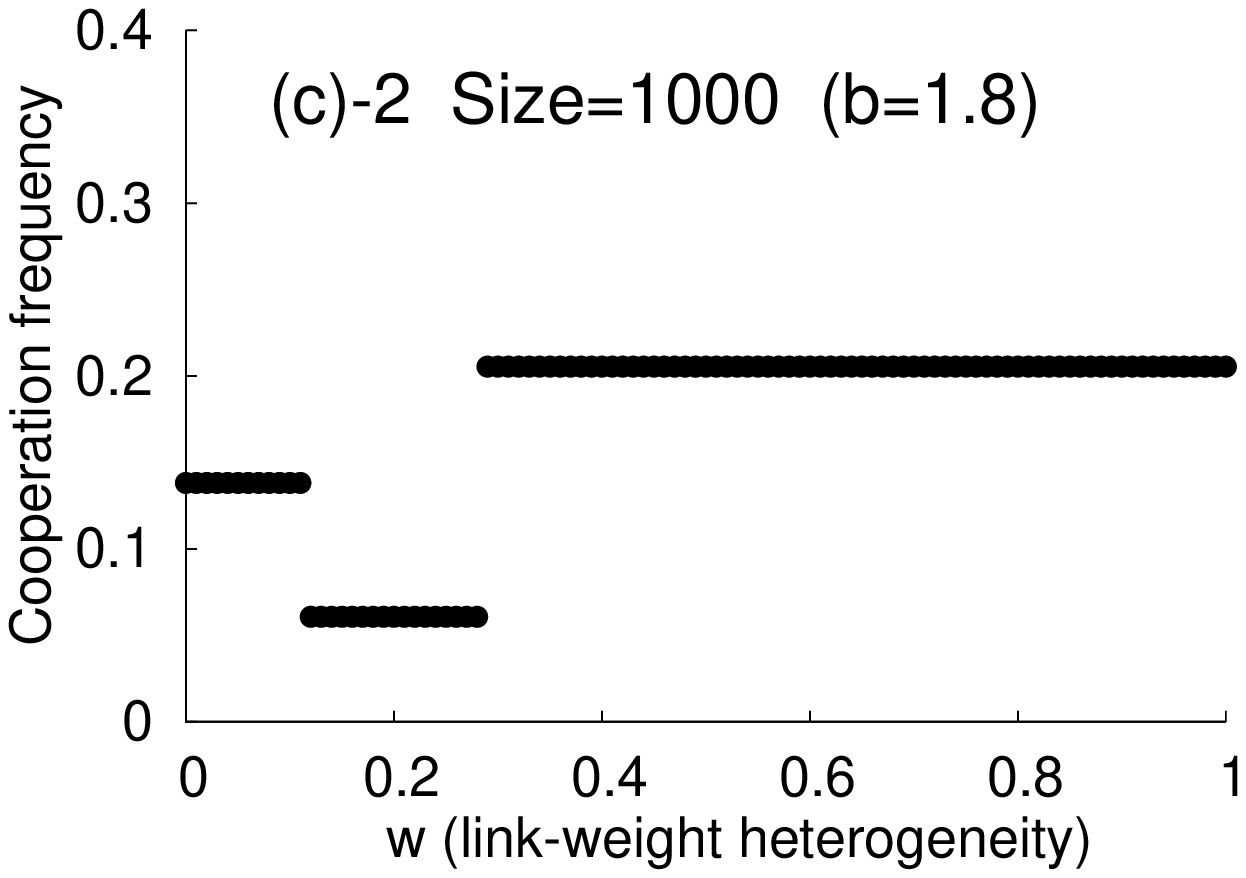}
\includegraphics[scale=0.31]{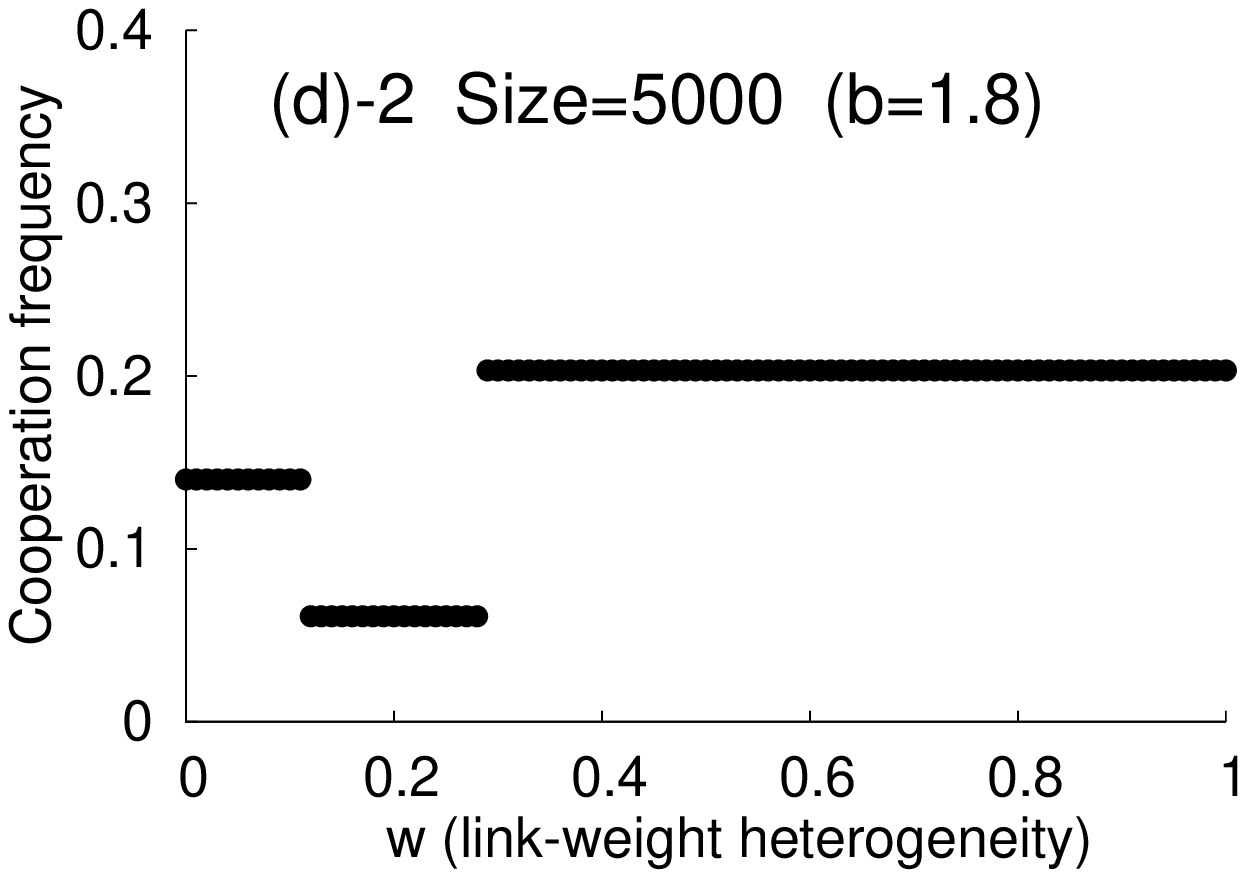}\hspace{3.5pt}\includegraphics[scale=0.31]{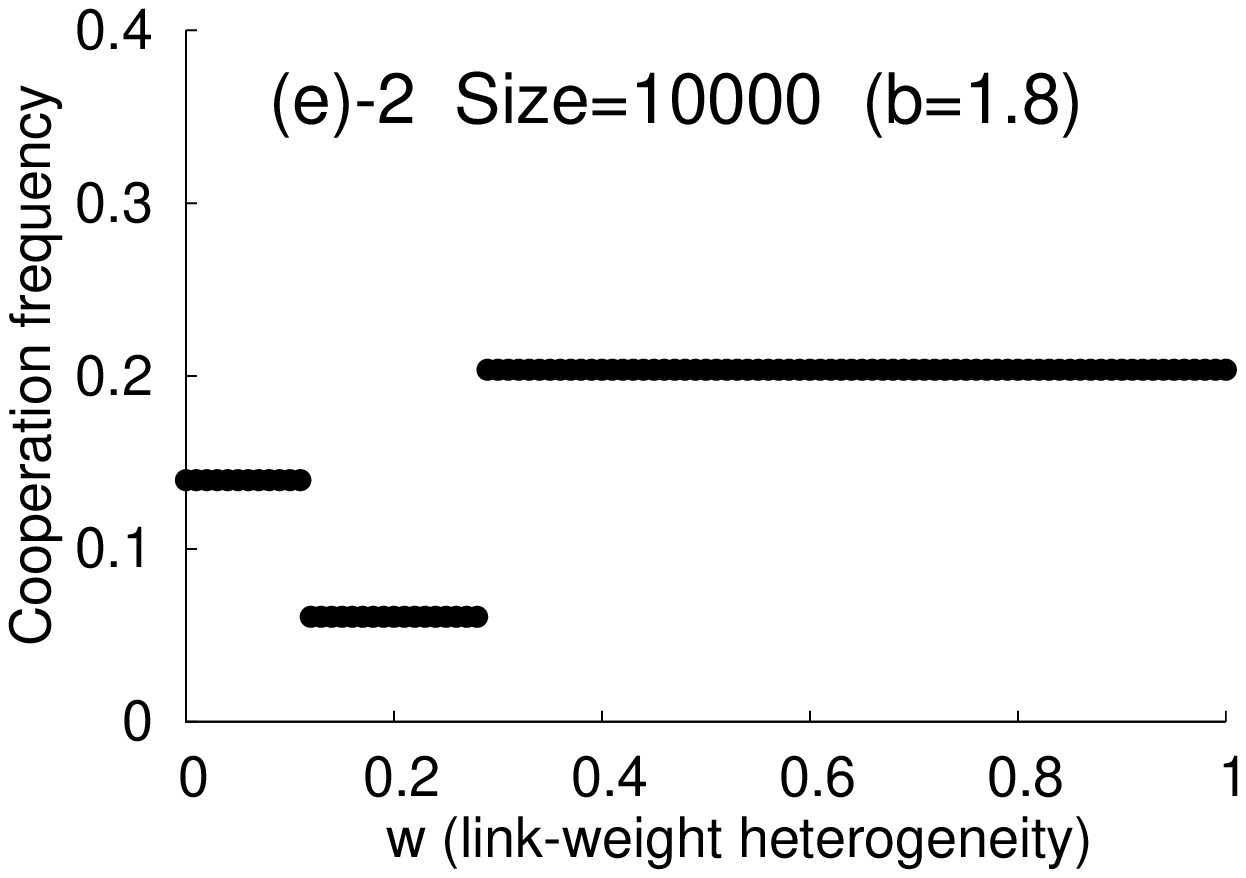}
\caption{Frequency of cooperation for different values of link-weight heterogeneity, $w$, in a weighted one-dimensional lattice: (a)-1 and (a)-2 for size 100, (b)-1 and (b)-2 for size 500, (c)-1 and (c)-2 for size 1,000, (d)-1 and (d)-2 for size 5,000, and (e)-1 and (e)-2 for size 10,000.
(a) to (e) -1 represent the results with a small $b$ ($b$=1.2), and (a) to (e) -2 show the results with a large $b$ ($b$=1.8).The horizontal axis represents the degree of $w$, which reflects the magnitude of link-weight heterogeneity.
The vertical axis indicates the cooperation frequency.}
\label{Robustness against the system size}
\end{figure}

Fig.~\ref{Robustness against the system size} shows the cooperation frequency for different values of link-weight heterogeneity $w$.
Figs.~\ref{Robustness against the system size}(a)-1 to (e)-1 illustrate the simulation results for different system sizes when $b$, temptation to defect, is small ($b$=1.2).
Similarly, Figs.~\ref{Robustness against the system size}(a)-2 to (e)-2 illustrate the simulation results when $b$ is large ($b$=1.8).

From these figures, we can see that the results are not different with a variation in system size, as in the following.
(i) We observe that homogeneous link-weight heterogeneity ($w$=0) does not always yield the highest cooperation level in an entire range of $w$.
That is, the moderate magnitude of heterogeneity achieves more cooperation than the case in which the link weight is homogeneous.
(i\hspace{-.1em}i) We observe that there are some thresholds in $w$ at which discontinuous changes of the cooperation frequency occur.
These tendencies can be seen in the aforementioned six one-dimensional lattice networks.
Therefore, we believe that the results obtained are robust with regard to differences in the system size.

\subsection{Robustness of the results against the mutation/error}

To check the robustness of the results against the mutation/error, we built a model that incorporates the possibility of a decision-making error.
We performed a computer simulation in which each individual made a mistake in the process of decision-making for a strategy update with probability 0.02, 0.05, 0.08, and 0.1, and we compared the results with that of the case without a decision-making error.

\begin{figure}[htbp]
%%\centering
\includegraphics[scale=0.31]{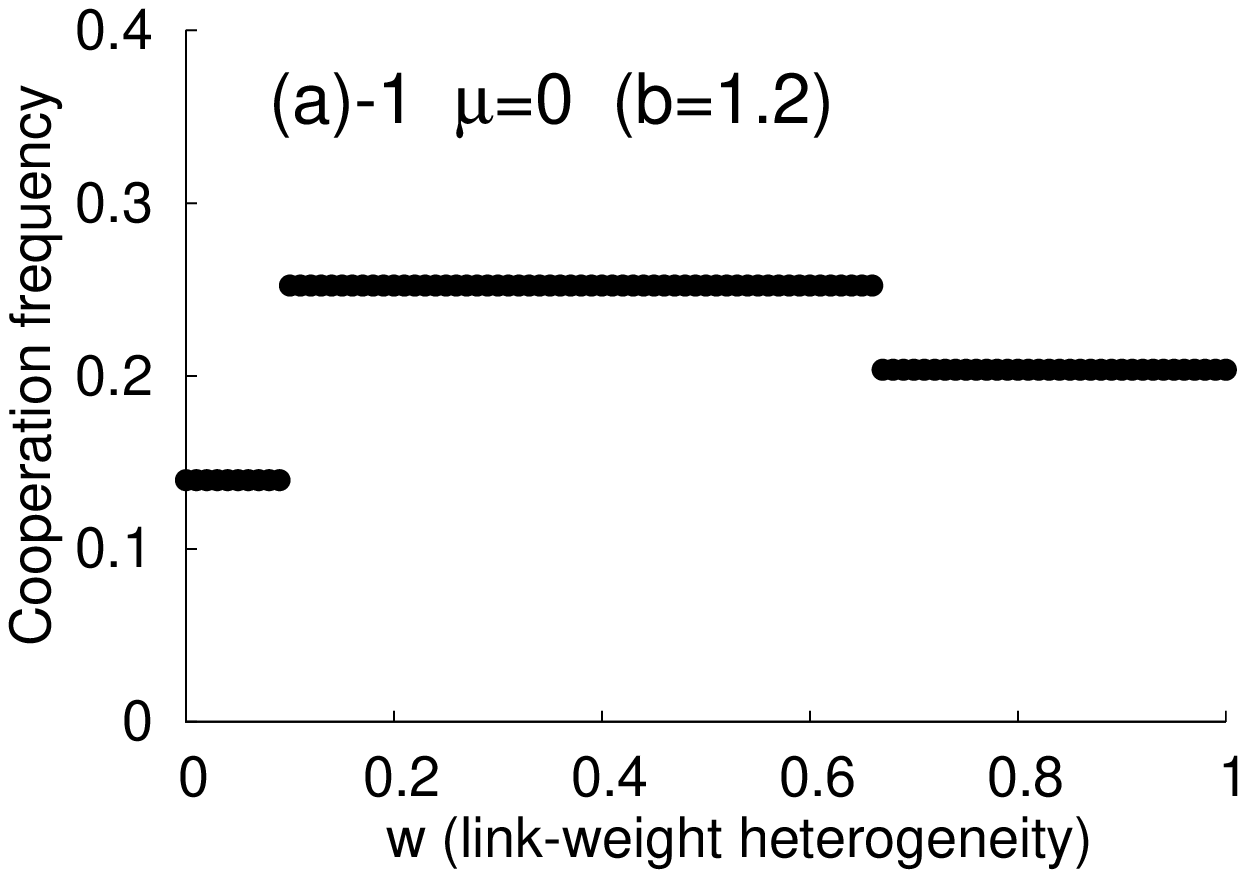}\hspace{3.5pt}\includegraphics[scale=0.31]{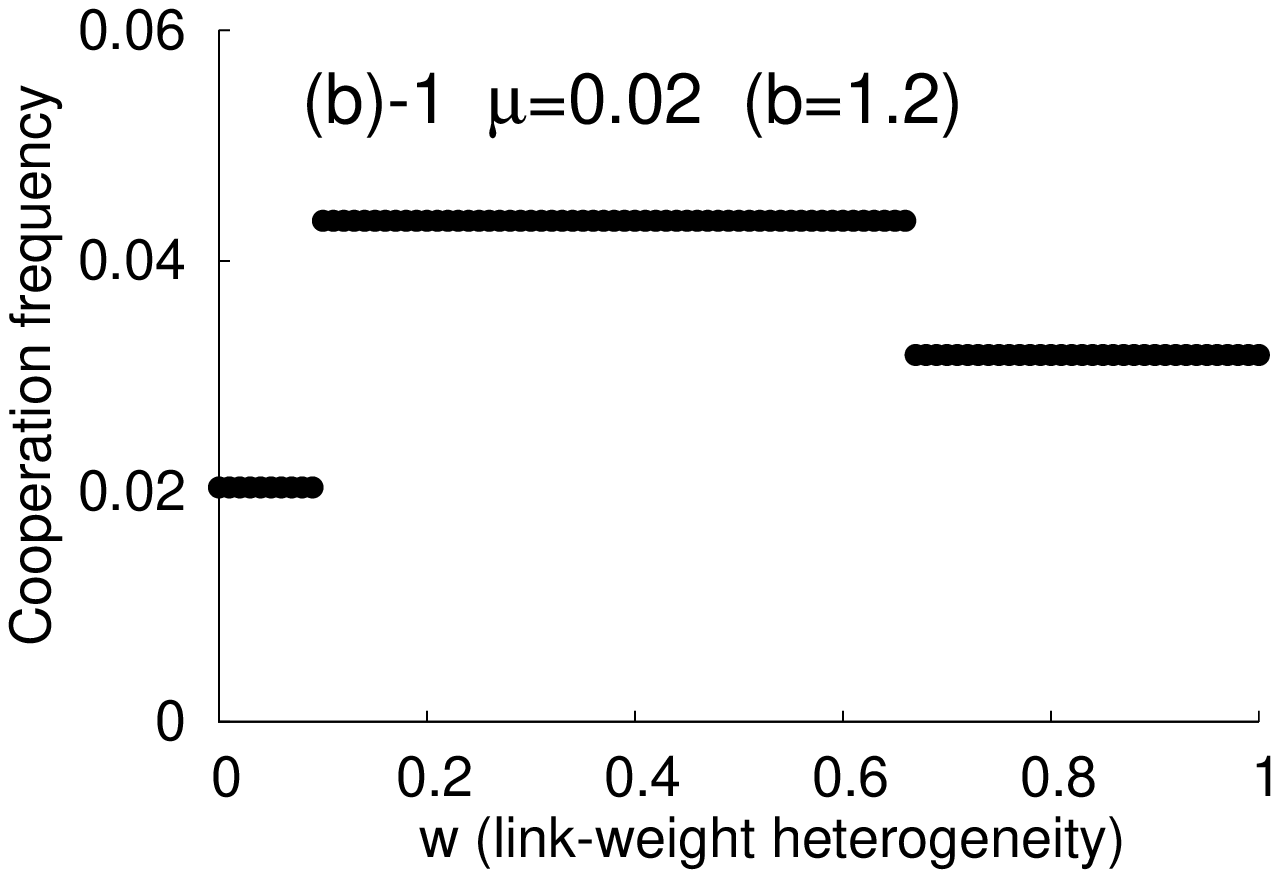}\hspace{3.5pt}\includegraphics[scale=0.31]{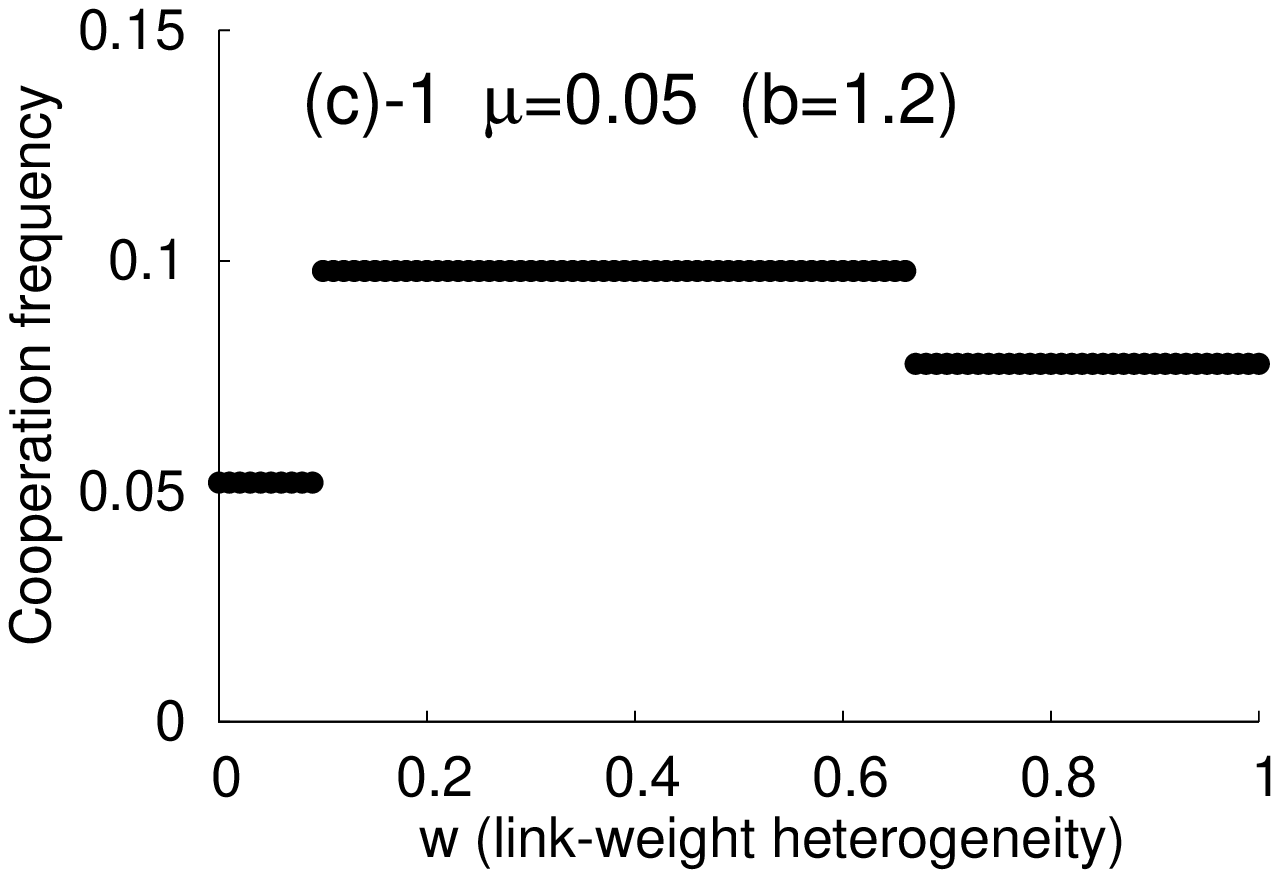}
\includegraphics[scale=0.31]{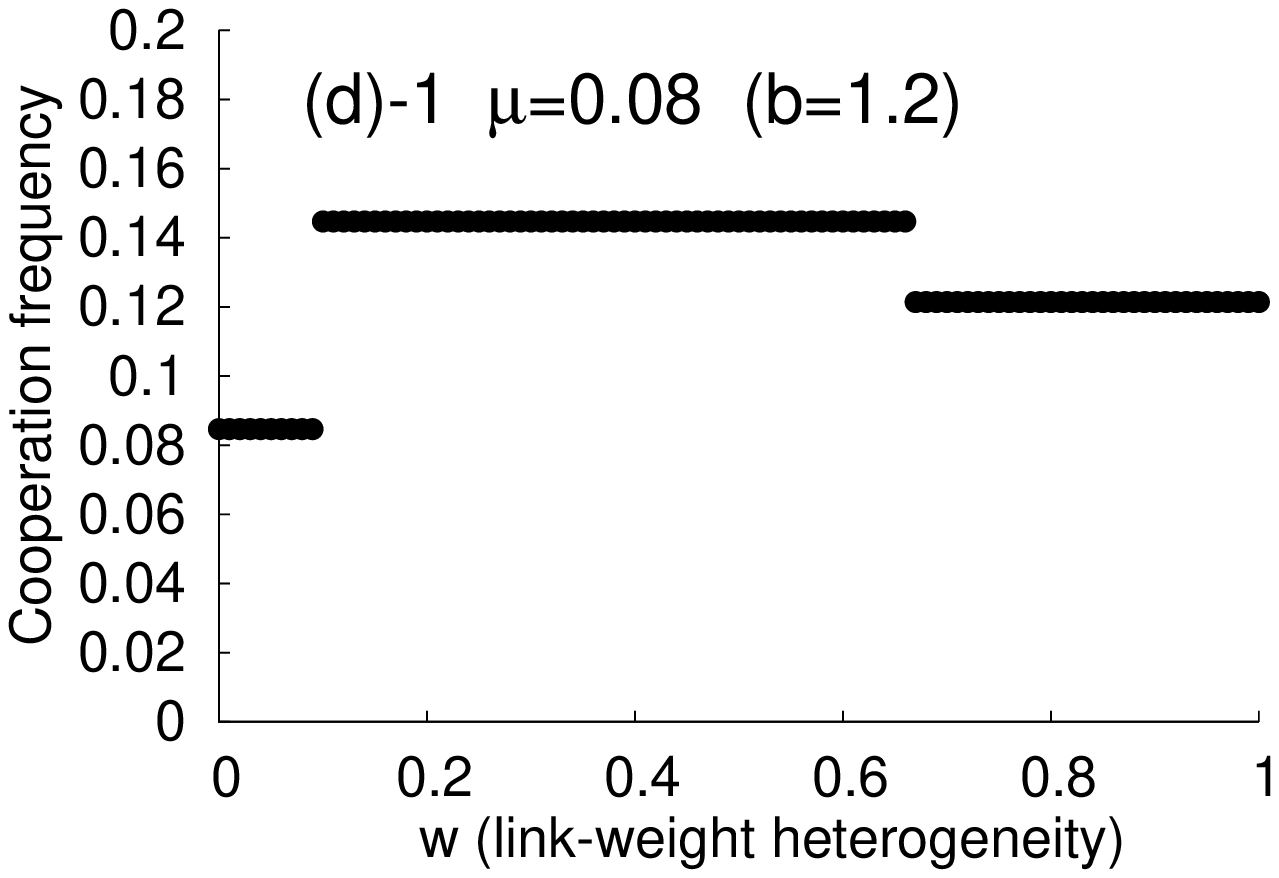}\hspace{3.5pt}\includegraphics[scale=0.31]{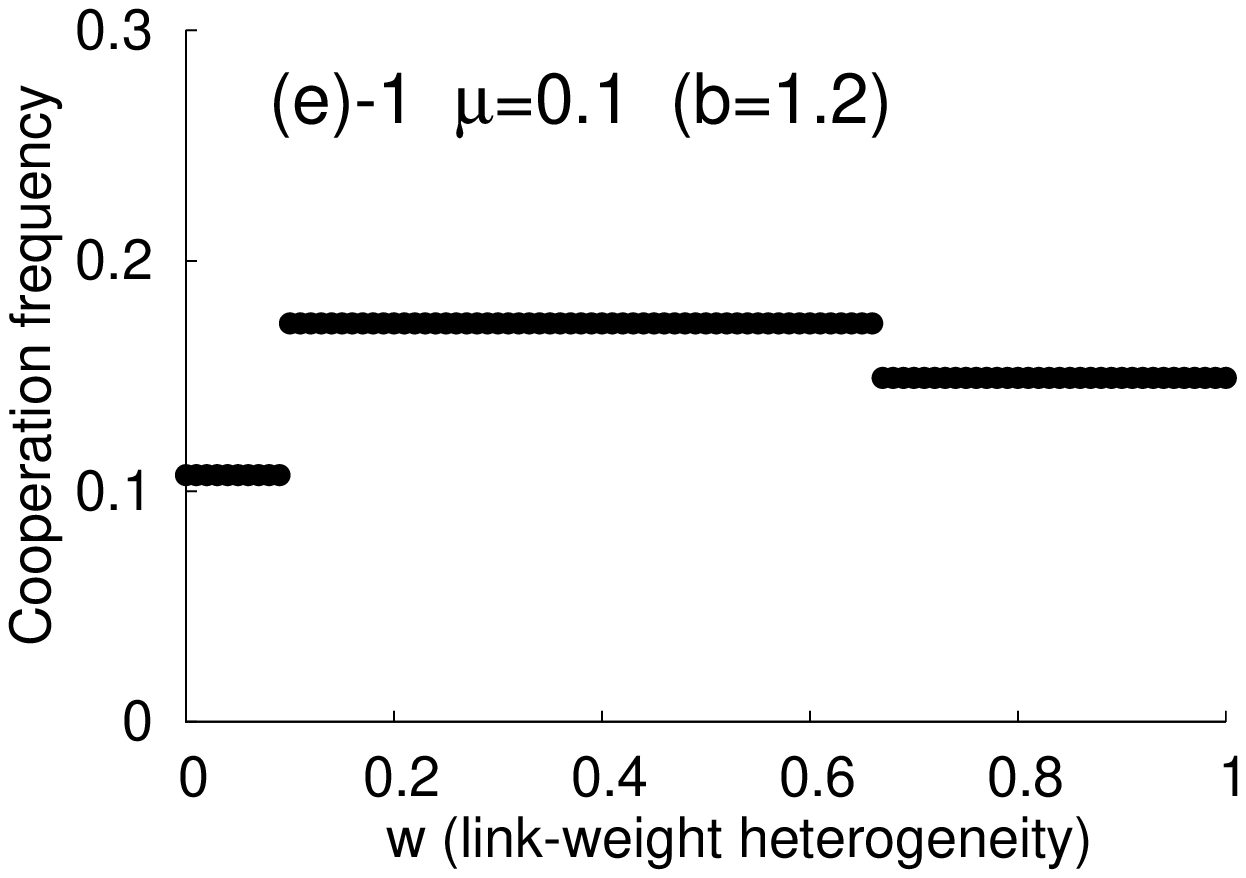}

\includegraphics[scale=0.31]{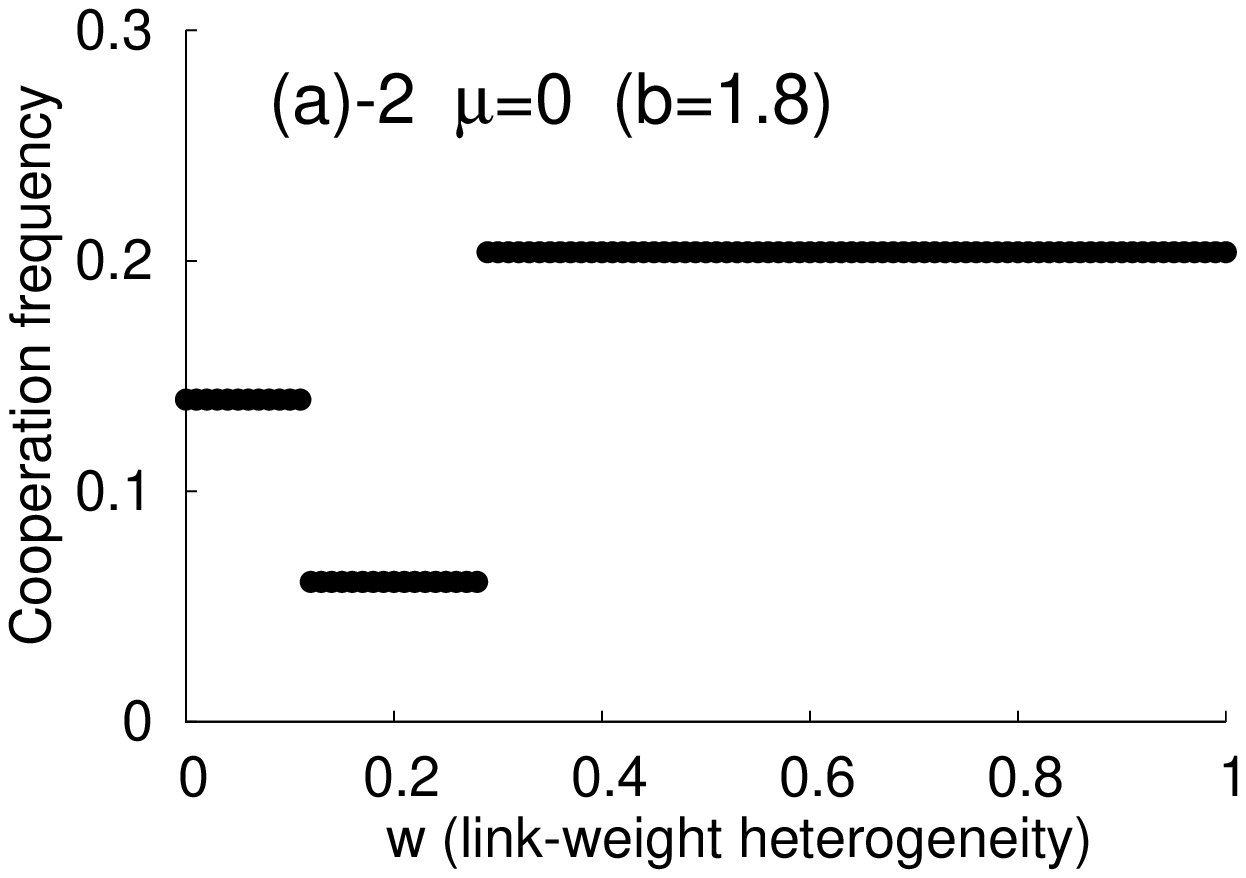}\hspace{3.5pt}\includegraphics[scale=0.31]{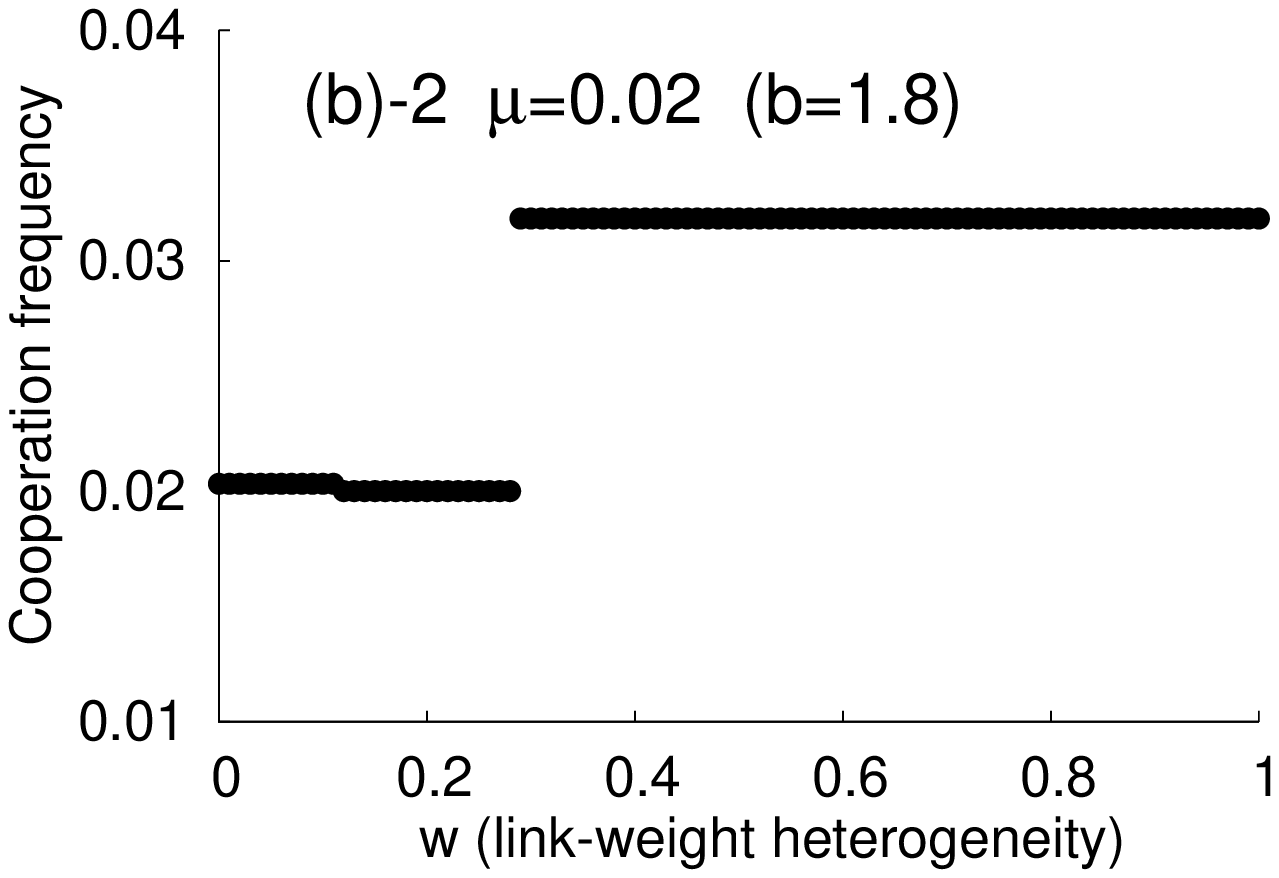}\hspace{3.5pt}\includegraphics[scale=0.31]{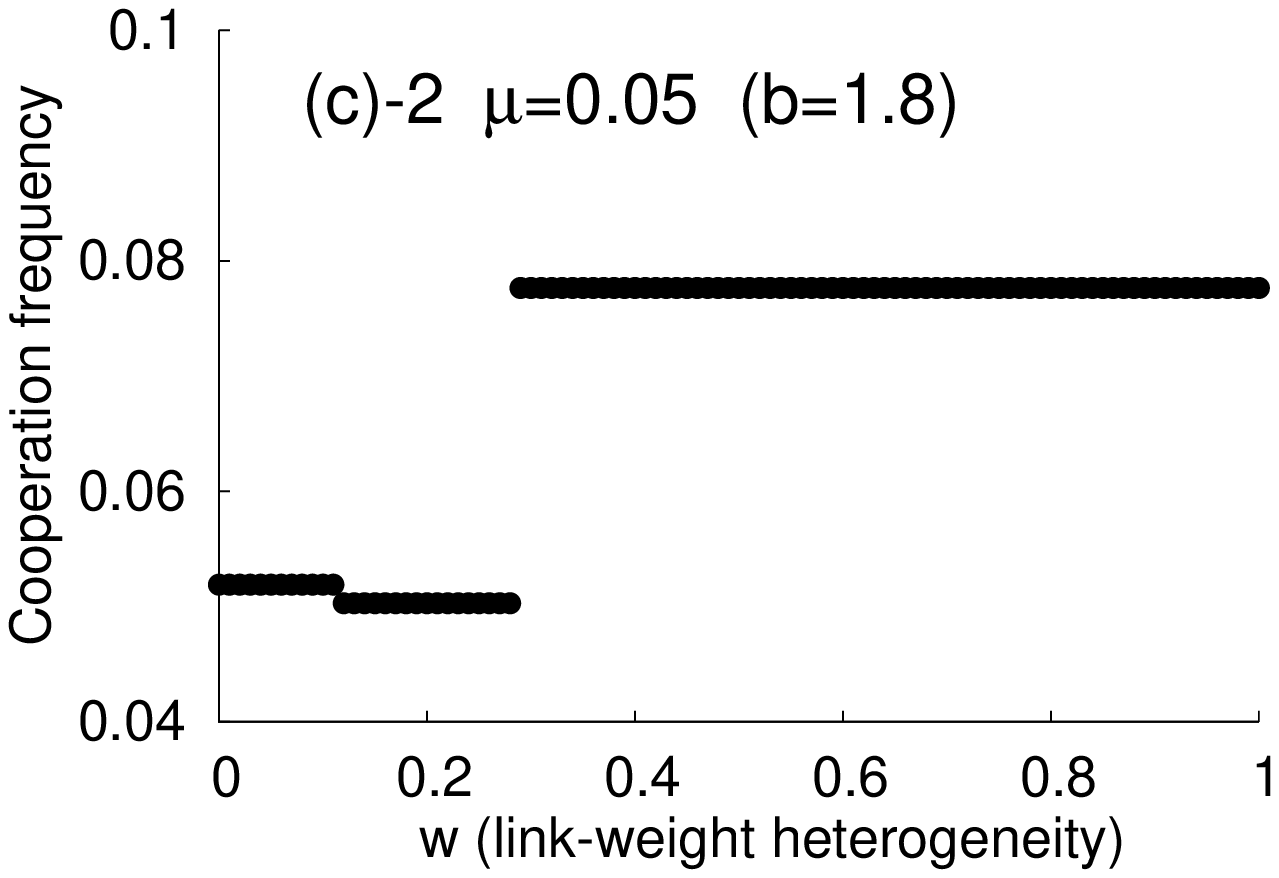}
\includegraphics[scale=0.31]{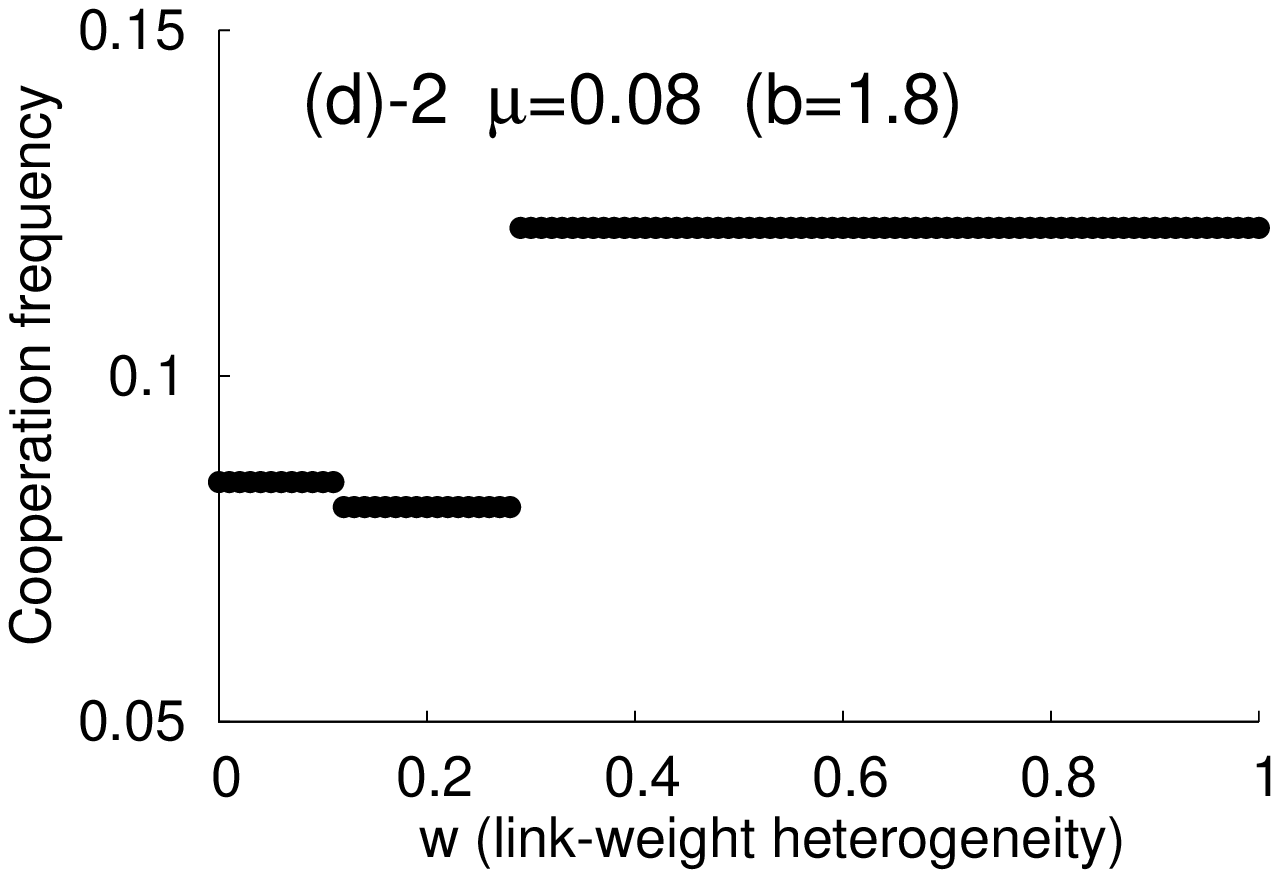}\hspace{3.5pt}\includegraphics[scale=0.31]{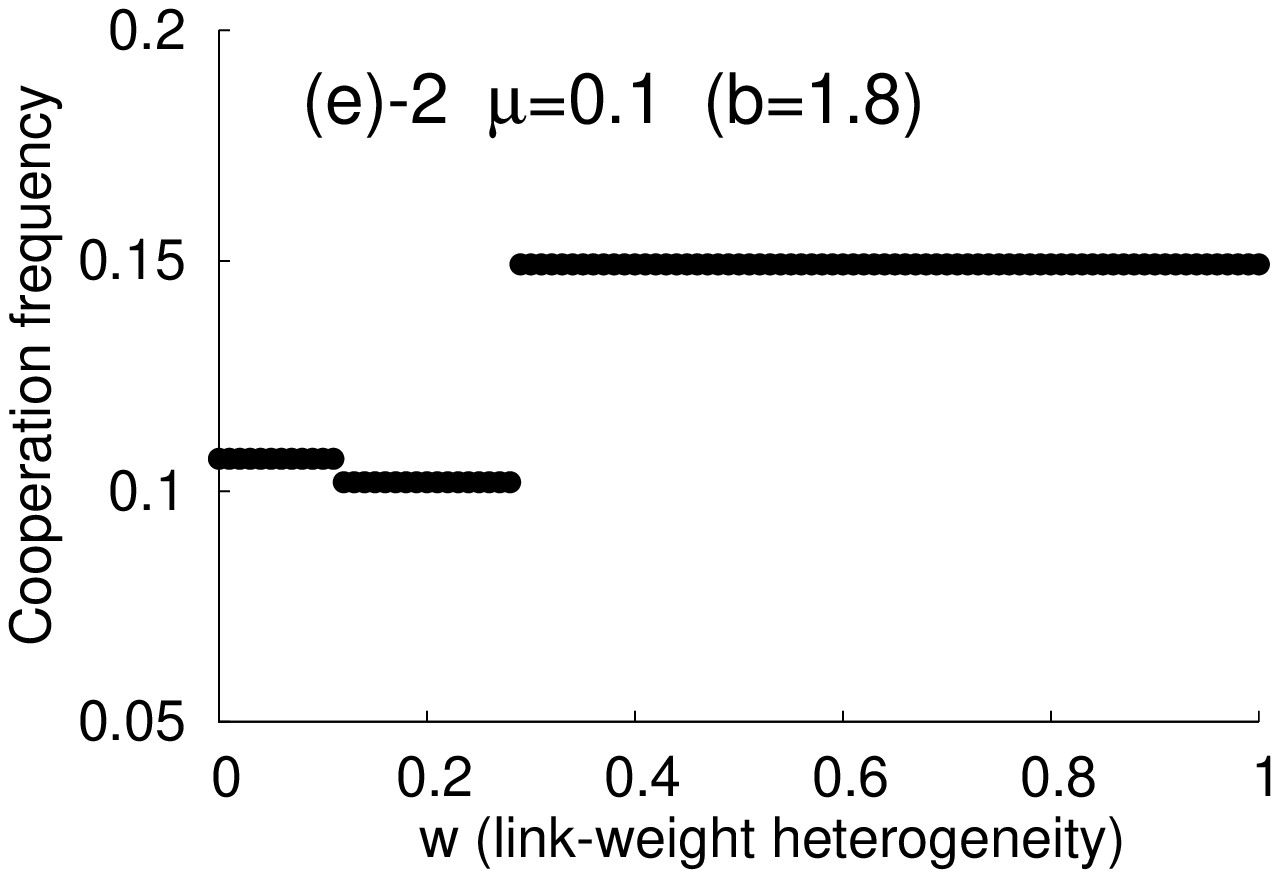}
\caption{Frequency of cooperation for different values of link-weight heterogeneity, w, in a weighted one-dimensional lattice: (a) error rate 0.02, (b) error rate 0.05, (c) error rate 0.08, and (d) error rate 0.1.
(a) to (e) -1 represent the results with a small $b$ ($b$=1.2), and (a) to (e) -2 show the results with a large $b$ ($b$=1.8).
The horizontal axis represents the degree of $w$, which reflects the magnitude of link-weight heterogeneity.
The vertical axis indicates the cooperation frequency.}
\label{Robustness against the error}
\end{figure}

Fig.~\ref{Robustness against the error} shows the cooperation frequency for different values of link-weight heterogeneity $w$.
Figs.~\ref{Robustness against the error}(a)-1 to (e)-1 illustrate the simulation results for different error rates when $b$, the temptation to defect from a cooperator, is small ($b$=1.2).
Similarly, Figs.~\ref{Robustness against the error}(a)-2 to (e)-2 illustrate the simulation results when $b$ is large ($b$=1.8).

From these figures, we can see that the results are not different regardless of the error rate, as in the following.
(i) We observe that homogeneous link-weight heterogeneity ($w$=0) does not always yield the highest cooperation level in an entire range of $w$.
That is, the moderate magnitude of heterogeneity achieves more cooperation than the case in which the link weight is homogeneous.
(i\hspace{-.1em}i) We observe that there are some thresholds in $w$ at which discontinuous changes of the cooperation frequency occur.
These tendencies can be seen in the aforementioned six one-dimensional lattice networks.
Although an error may cause the birth of a defective individual, it is likely that \emph{strategy configuration and condition} for the spread or maintenance of cooperation can prevent the penetration of defection in the population.
Thus, we confirmed that our results hold true even when there is an error in the decision making of the strategy update of each individual.

\section{Classification of strategy configurations}\label{Appendix}

In this appendix, we show the classification of the initial strategy configurations for a convergent state of cooperation frequency in a small one-dimensional lattice consisting of six individuals.
Of the $2^{6}$=64 initial strategy configurations, we focus on those configurations that, through evolution, reach different cooperative states depending on whether the link weight is heterogeneous ($w>0$) or homogeneous ($w=0$) as a result of the evolution of strategies.
As mentioned in Section~\ref{subsection: Analysis of small population case}, the initial strategy configurations are classified into the following three types: Type (i) in which heterogeneous link weight ($w>0$) leads to higher cooperation frequency than the homogeneous one ($w=0$); Type (i\hspace{-.1em}i) in which heterogeneous weight suppresses cooperation; and Type (i\hspace{-.1em}i\hspace{-.1em}i) where both heterogeneous and homogeneous link weights lead to the same magnitude of cooperation. 
The initial strategy configuration types are listed in Table~\ref{A list of initial strategy configurations and classification}\footnote{Six strategy configurations are classified as both Type (i) and (i\hspace{-.1em}i), where, whether the heterogeneous link weight ($w>0$) achieves a higher cooperation frequency through evolution than the homogeneous weight ($w=0$) depends on the value of $b$.
Our purpose was to identify the mechanisms whereby link-weight heterogeneity enhances cooperation and to derive the conditions for the mechanisms to work.
Therefore, we investigated the strategy configurations classified only as Type (i) or (i\hspace{-.1em}i).}.

\setcounter{table}{0}
\begin{table}[htbp]
\centering
\scalebox{0.9}{
\begin{tabular}{ll} \hline
Initial strategy configuration & Classification type\\ \hline
$-$C$\equiv $C$-$D$\equiv $D$-$C$\equiv $C$-$, $-$D$\equiv $D$-$C$\equiv $C$-$C$\equiv $C$-$,& Type (i): Heterogeneous link\\
$-$C$\equiv $C$-$C$\equiv $C$-$D$\equiv $D$-$, $-$C$\equiv $C$-$D$\equiv $D$-$D$\equiv $D$-$,& weight ($w>0$) promotes\\
$-$D$\equiv $D$-$D$\equiv $D$-$C$\equiv $C$-$, $-$D$\equiv $D$-$C$\equiv $C$-$D$\equiv $D$-$& further cooperation\\ \hline
$-$C$\equiv $C$-$C$\equiv $D$-$D$\equiv $C$-$, $-$C$\equiv $D$-$D$\equiv $C$-$C$\equiv $C$-$,& Type (i\hspace{-.1em}i): Homogeneous link\\
$-$D$\equiv $C$-$C$\equiv $C$-$C$\equiv $D$-$& weight ($w=0$) promotes\\
& further cooperation\\ \hline
$-$C$\equiv $C$-$C$\equiv $C$-$C$\equiv $C$-$, $-$C$\equiv $C$-$C$\equiv $D$-$C$\equiv $D$-$,& Type (i\hspace{-.1em}i\hspace{-.1em}i): Heterogeneous\\
$-$C$\equiv $C$-$D$\equiv $C$-$D$\equiv $C$-$, $-$C$\equiv $C$-$D$\equiv $C$-$C$\equiv $D$-$,& weight ($w>0$) and\\
$-$C$\equiv $D$-$C$\equiv $C$-$D$\equiv $C$-$, $-$C$\equiv $D$-$C$\equiv $D$-$C$\equiv $C$-$,& homogeneous weight\\
$-$C$\equiv $D$-$C$\equiv $D$-$D$\equiv $C$-$, $-$C$\equiv $D$-$C$\equiv $C$-$C$\equiv $D$-$,& ($w=0$) achieve the\\
$-$C$\equiv $D$-$C$\equiv $D$-$C$\equiv $D$-$, $-$C$\equiv $D$-$C$\equiv $D$-$D$\equiv $D$-$,& same level of cooperation\\
$-$C$\equiv $D$-$D$\equiv $C$-$D$\equiv $C$-$, $-$C$\equiv $D$-$D$\equiv $D$-$D$\equiv $C$-$,&\\
$-$C$\equiv $D$-$D$\equiv $C$-$C$\equiv $D$-$, $-$C$\equiv $D$-$D$\equiv $C$-$D$\equiv $D$-$,&\\
$-$C$\equiv $D$-$D$\equiv $D$-$C$\equiv $D$-$, $-$C$\equiv $D$-$D$\equiv $D$-$D$\equiv $D$-$,&\\
$-$D$\equiv $C$-$C$\equiv $C$-$D$\equiv $C$-$, $-$D$\equiv $C$-$C$\equiv $D$-$C$\equiv $C$-$,&\\
$-$D$\equiv $C$-$C$\equiv $D$-$C$\equiv $D$-$, $-$D$\equiv $C$-$C$\equiv $D$-$D$\equiv $C$-$,&\\
$-$D$\equiv $C$-$C$\equiv $D$-$D$\equiv $D$-$, $-$D$\equiv $C$-$D$\equiv $C$-$C$\equiv $C$-$,&\\
$-$D$\equiv $C$-$D$\equiv $C$-$D$\equiv $C$-$, $-$D$\equiv $C$-$D$\equiv $C$-$C$\equiv $D$-$,&\\
$-$D$\equiv $C$-$D$\equiv $C$-$D$\equiv $D$-$, $-$D$\equiv $C$-$D$\equiv $D$-$C$\equiv $D$-$,&\\
$-$D$\equiv $C$-$D$\equiv $D$-$D$\equiv $C$-$, $-$D$\equiv $C$-$D$\equiv $D$-$D$\equiv $D$-$,&\\
$-$D$\equiv $D$-$C$\equiv $D$-$C$\equiv $D$-$, $-$D$\equiv $D$-$C$\equiv $D$-$D$\equiv $C$-$,&\\
$-$D$\equiv $D$-$C$\equiv $D$-$D$\equiv $D$-$, $-$D$\equiv $D$-$D$\equiv $C$-$D$\equiv $C$-$,&\\
$-$D$\equiv $D$-$D$\equiv $D$-$D$\equiv $C$-$, $-$D$\equiv $D$-$D$\equiv $C$-$C$\equiv $D$-$,&\\
$-$D$\equiv $D$-$D$\equiv $C$-$D$\equiv $D$-$, $-$D$\equiv $D$-$D$\equiv $D$-$C$\equiv $D$-$,&\\
$-$D$\equiv $D$-$D$\equiv $D$-$D$\equiv $D$-$&\\ \hline
\end{tabular}
}
\caption{List of initial strategy configurations and classification thereof into three types. 
The first row represents the initial strategy configurations; the second row shows the three classifications of the configurations in which greater link-weight heterogeneity $w$ promotes more cooperation, higher $w$ reduces cooperation frequency, and different values of $w$ do not have any effect on the magnitude of the cooperation level, respectively. 
In the table, ``C'' and ``D'' denote cooperator and defector, respectively, while ``$\equiv$'' indicates a link with a large weight, and ``$-$'' means a link with a small weight.}
\label{A list of initial strategy configurations and classification}
\end{table}

\section{Two-dimensional lattice: Results and brief analysis}\label{subsection: Results and brief analysis for a two-dimensional lattice}

Here we investigate the effect of link-weight heterogeneity on the evolution of cooperation on a two-dimensional lattice network.
We calculate the cooperation frequency and describe the brief analysis of the mechanism that heterogeneity promotes cooperation.
In our model, individuals are placed on the nodes of a two-dimensional lattice and are assumed to interact with their immediate neighbors.
We assume the periodic boundary condition for the network.
Each of them is surrounded by upper, lower, left, and right immediate neighbors and has two links with weight $w_{1}=1.0+w$ and two with weight $w_{2}=1.0-w$.
Thus we can control a value of link weight only by one parameter $w$ and link-weight amount of each individual can be 4.0 regardless of the value of $w$.

Under these assumptions, there can be infinitely many kinds of weighted two-dimensional lattices.
We show examples of these in Fig.~\ref{Examples of link weight distribution patterns}.
\setcounter{figure}{0}
\begin{figure}[h]
\centering
\includegraphics[scale=0.24]{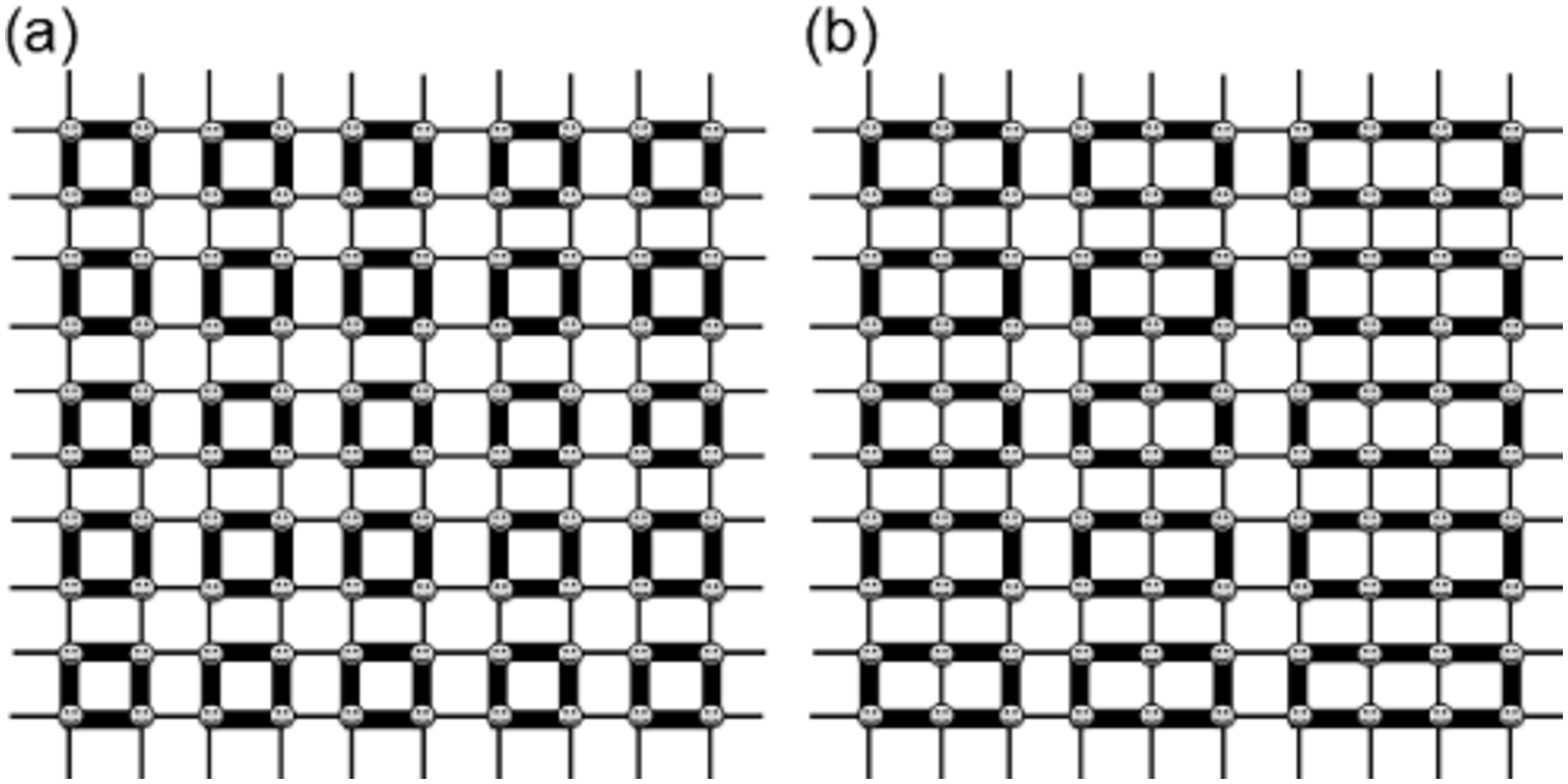}\includegraphics[scale=0.24]{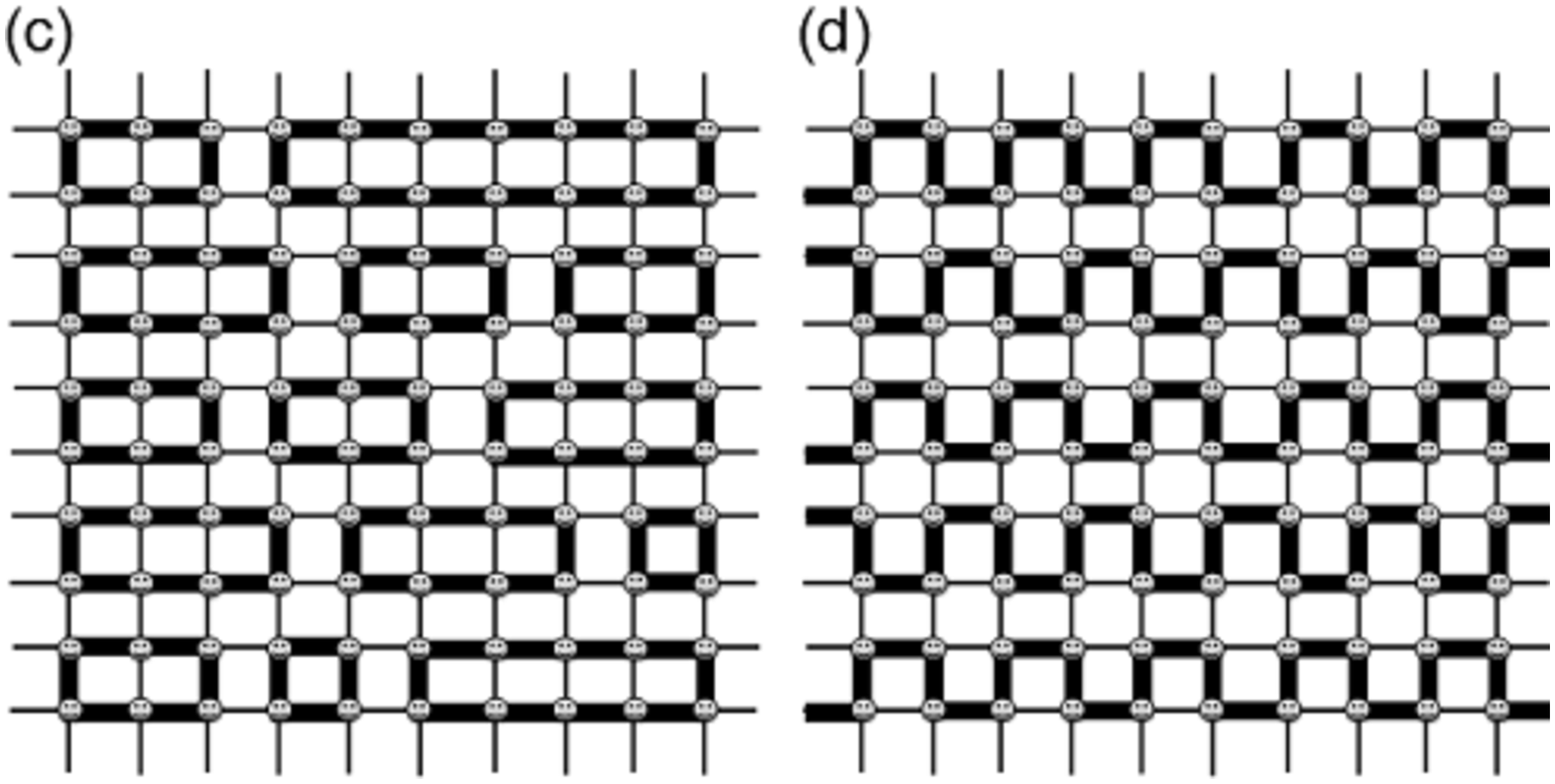}
\includegraphics[scale=0.24]{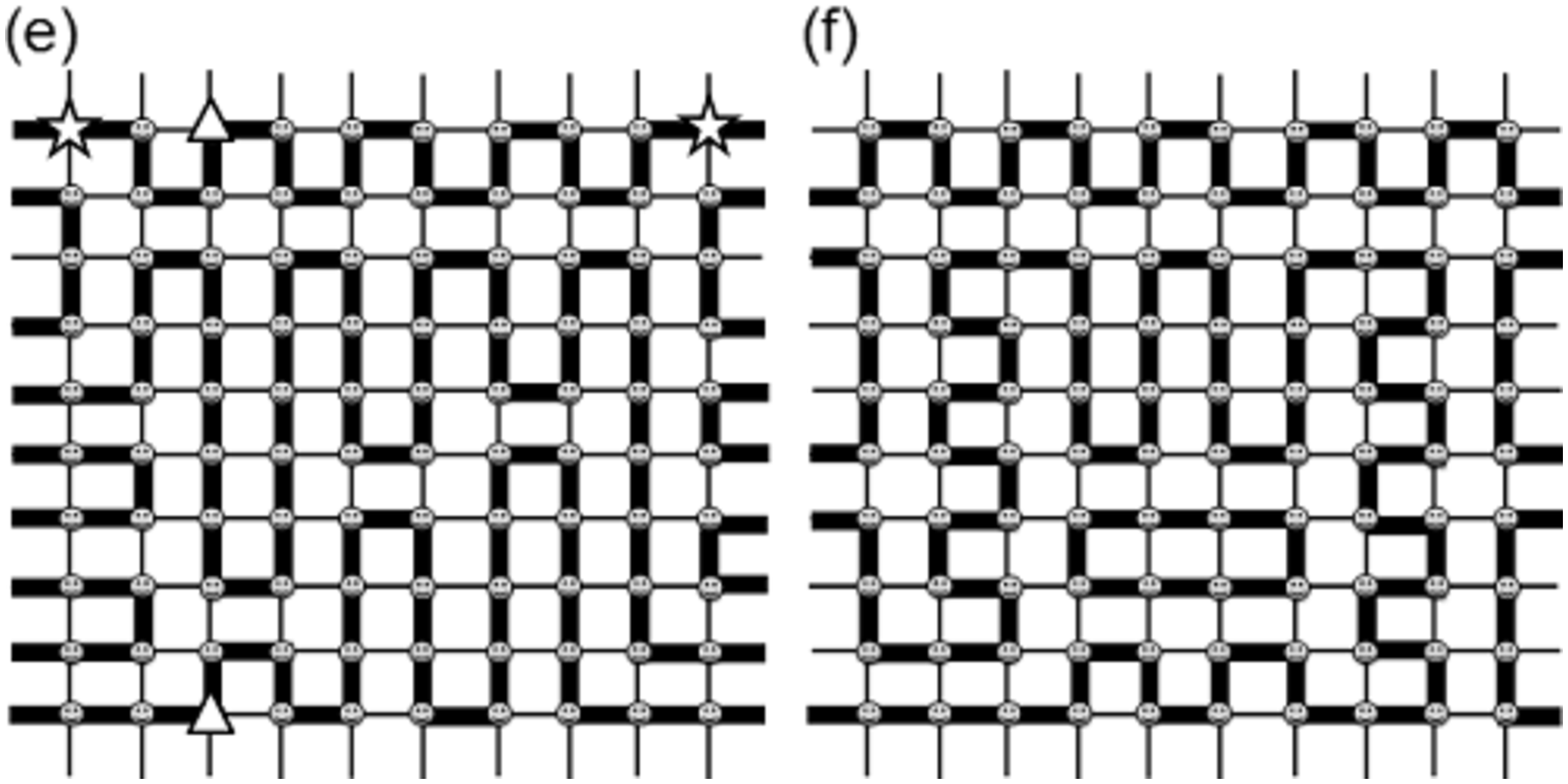}\includegraphics[scale=0.24]{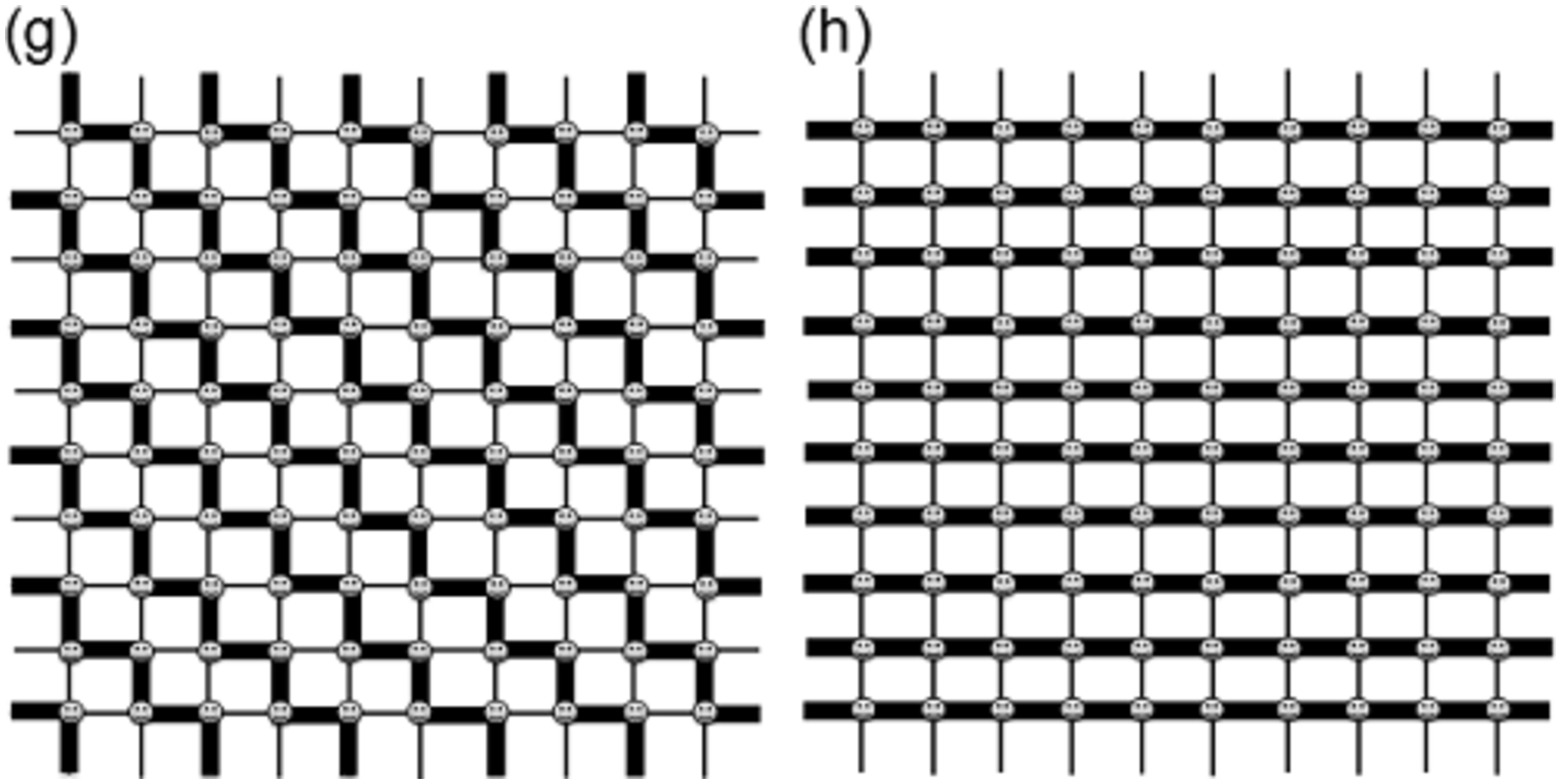}
\caption{Examples of link weight distribution patterns in a two-dimensional lattice without \emph{inter-individual heterogeneity} and where each individual has two links with a large weight (denoted by thick lines) and two links with a small weight (denoted by thin lines).
Here, (a), (b), and (c) show weighted lattice networks filled with non-overlapping rectangles composed of large-weight links, while (d), (e), (f), (g), and (h) show those filled with non-overlapping closed lines forming non-rectangular shapes.
Because the network is assumed to satisfy periodic boundary conditions, the individuals marked with stars on the right and left sides of (e) are connected with one another.
Similarly, the individuals marked with triangles at the top and bottom are connected with one another.}
\label{Examples of link weight distribution patterns}
\end{figure}
As shown in Fig.~\ref{Examples of link weight distribution patterns}(a), (b), and (c), when a weighted lattice is filled with non-overlapping rectangles composed of large-weight links, each individual (node) is either a vertex (corner) of the rectangle or on a side of the rectangle.
Observation of these figures enables us to understand easily that there can be infinitely many link-weight distribution patterns of a two dimensional lattice occupied by such examples.
However, Fig.~\ref{Examples of link weight distribution patterns}(d), (e), (f), (g), and (h) shows examples filled with non-overlapping closed lines that form non-rectangular shapes.
Our analysis of the evolution of cooperation in a two-dimensional lattice focused on the cases in Fig.~\ref{Examples of link weight distribution patterns}(g) and (h).
We call the former \emph{Pattern A} and the latter \emph{Pattern B} in this paper.

\begin{figure}[htbp]
\centering
\includegraphics[scale=0.22]{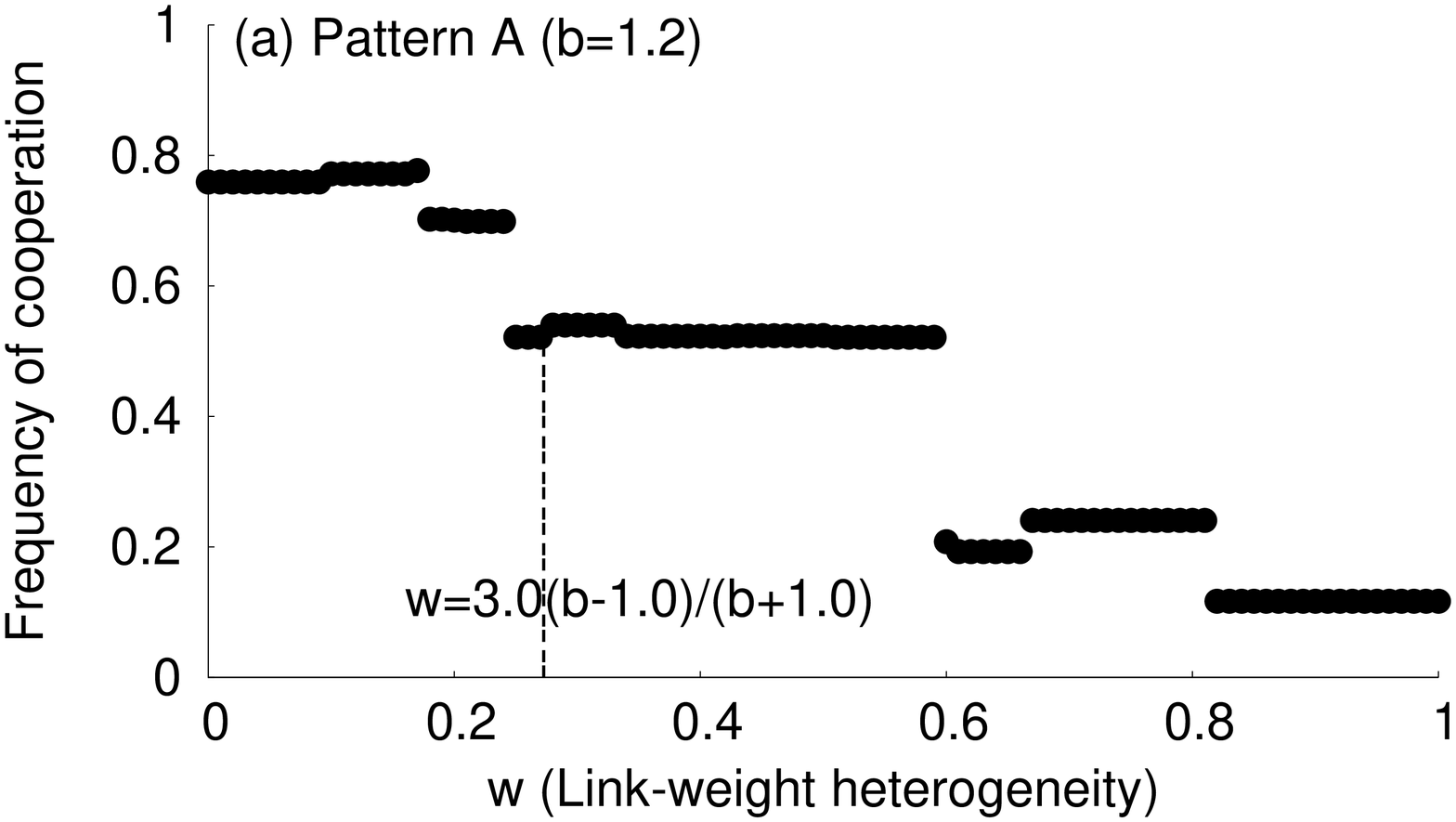}\includegraphics[scale=0.22]{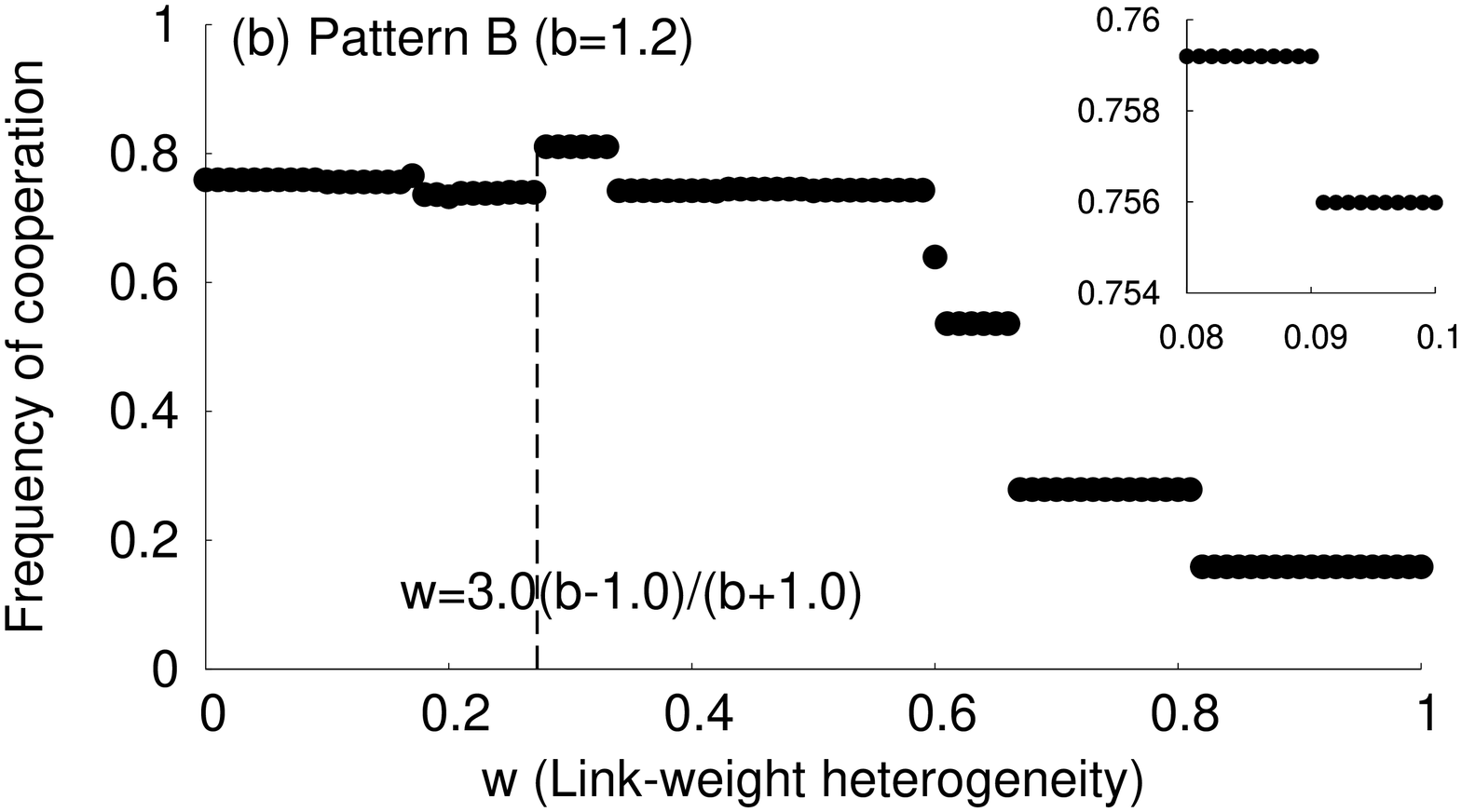}
\caption{Frequency of cooperation for different values of link-weight heterogeneity $w$ obtained through simulations of the PD game on a weighted two-dimensional lattice.
The horizontal and vertical axes denote the degree of link-weight heterogeneity $w$ and cooperation frequency, respectively:
(a) results for a \emph{pattern A} network structure, and (b) results for \emph{pattern B}.
Both experiments used $b=1.2$.}
\label{Cooperation frequency on a two-dimensional lattice}
\end{figure}

Figs.~\ref{Cooperation frequency on a two-dimensional lattice}(a) and (b) give the simulation results for the PD game on a weighted two-dimensional lattice.
These graphs show the frequency of cooperation for different values of link-weight heterogeneity $w$.
Fig.~\ref{Cooperation frequency on a two-dimensional lattice}(a) shows the simulation results when the population network follows \emph{pattern A}, while Fig.~\ref{Cooperation frequency on a two-dimensional lattice}(b) shows the same for \emph{pattern B}, both of which set $b=1.2$.

Contrary to the case of a one-dimensional lattice, it may be difficult to state that large heterogeneity of link weight promotes further cooperation.
However, as seen in Fig.~\ref{Cooperation frequency on a two-dimensional lattice}(a) and (b), link-weight heterogeneity $w$ has a certain heterogeneous value, which enhances further cooperation than in the case of homogeneous weight.
As in the case of a one-dimensional lattice, we observe that cooperation frequency changes in a stepwise manner with an increase in link-weight heterogeneity $w$.

In a two-dimensional lattice, the number of interactions of each individual is twice that in the case of a one-dimensional lattice, so there may be a great number of conditions under which a difference in the degree of link-weight heterogeneity yields a different level of cooperation frequency.
The vast number of thresholds makes it difficult to determine whether cooperation frequency changes in a stepwise manner with a change in the value of link-weight heterogeneity $w$.
However, as shown in Fig.~\ref{Cooperation frequency on a two-dimensional lattice}(a) and (b), and the enlarged figure in the top right corner of Fig.~\ref{Cooperation frequency on a two-dimensional lattice}(b), it is highly likely that in the two-dimensional lattice, the cooperation frequency changes, as in the one-dimensional lattice, in a stepwise manner at the threshold values for link-weight heterogeneity $w$.
Although we have not yet identified all the threshold values of $w$ at which the frequency of cooperation changes in a stepwise manner for a given value of $b$, we can present an example of \emph{the condition for the spread of cooperation}, $w>3.0(b-1.0)/(b+1.0)$.

\begin{figure}[htbp]
\centering
\includegraphics[scale=0.25]{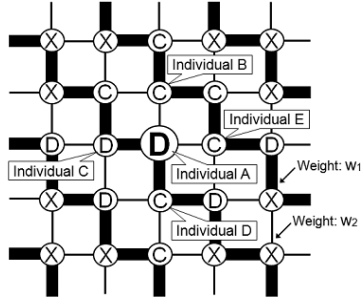}
\caption{Example of the strategy configuration patterns in a two-dimensional lattice with \emph{Pattern A}. 
We focus on \emph{Individual A} as the \emph{focal individual} and on \emph{Individuals B, C, D, and E} who are neighbors of \emph{Individual A}.
In this figure, ``C'' and ``D'' denote cooperator and defector, respectively, and ``X'' represents an individual with an arbitrary strategy, ``C'' or ``D''.
Note that the strategies of individuals ``X'' do not affect the payoffs of \emph{Individual A} and \emph{Individual B}.}
\label{An example of the strategy configuration patterns on a two-dimensional lattice}
\end{figure}

To perform the similar analisys to that in the case of a one-dimensional lattice, we use a with 5$\times $5 sized small lattice and focus on the dynamics of the strategy of one individual.
We show an example of the strategy configuration patterns in a two-dimensional lattice with \emph{Pattern A} in Fig.~\ref{An example of the strategy configuration patterns on a two-dimensional lattice}.

We focus on \emph{Individual A} as the \emph{focal individual} and on his/her strategy update from defection to cooperation.
Moreover, we focus on \emph{Individuals B, C, D, and E}, neighbors of \emph{Individual A}.
If the evolutionary dynamics starts from the strategy configurations in this figure, \emph{Individual A} decides to cooperate in the next step and spreads the cooperative state if the link-weight heterogeneity satisfies the inequality given later. 
We explain in detail why this spread of cooperation occurs.

After playing the PD game, \emph{Individual A} gets the score of $b(3.0-w)$, and similarly the scores of \emph{Individuals B, C, D, and E} are $w+3.0$, $b(1.0+w)$, $1.0-w$, and $1.0+w$, respectively.
\emph{Individual A} imitates the strategy of \emph{Individual B} and changes his/her strategy from defection to cooperation (spread of cooperation) if the score of \emph{Individual B} is higher than that of \emph{Individual A}.
Thus, the condition under which \emph{Individual A} imitates the strategy of \emph{Individual B} is $b(3.0-w)<w+3.0$; that is, $w>3.0(b-1.0)/(b+1.0)$ for a given $b$.

As already mentioned, we observed a stepwise change of the cooperation frequency when $w$ is placed at the threshold $3.0(b-1.0)/(b+1.0)$.
In addition, we confirmed that the threshold $3.0(b-1.0)/(b+1.0)$ is one of the conditions for $w$ to achieve the spread of cooperation.
Because it is difficult to find all the \emph{spread/maintenance pattern strategy configurations} that reach different cooperative states depending on the value of link-weight heterogeneity $w$ and to derive all the \emph{conditions for the spread/maintenance of cooperation}, discovery of these for a two-dimensional lattice is left for future work.

\end{document}